\documentclass[fleqn,usenatbib]{mnras}

\usepackage[T1]{fontenc}

\DeclareRobustCommand{\VAN}[3]{#2}
\let\VANthebibliography\thebibliography
\def\thebibliography{\DeclareRobustCommand{\VAN}[3]{##3}\VANthebibliography}

\usepackage{graphicx}	
\usepackage{amsmath}	
\usepackage{amssymb}	

\usepackage{newtxtext,newtxmath}

\title[Observing simulated star cluster formation]{A panchromatic view of star cluster formation in a simulated dwarf galaxy starburst}

\author[Lah\'en et al.]{
Natalia Lah\'en$^{1}$\thanks{E-mail: nlahen@mpa-garching.mpg.de},
Thorsten Naab$^{1}$ and 
Guinevere Kauffmann$^{1}$
\\
$^{1}$Max Planck Institute for Astrophysics, Karl-Schwarzschild-Stra$\beta$e 1, D-85741 Garching, Germany\\
}

\date{Accepted 2022 June 06. Received 2022 June 05; in original form 2021 October 20}

\pubyear{2022}

\begin{document}
\label{firstpage}
\pagerange{\pageref{firstpage}--\pageref{lastpage}}
\maketitle

\begin{abstract}

We present a photometric analysis of star and star cluster (SC) formation in a high-resolution simulation of a dwarf galaxy starburst that allows the formation of individual stars to be followed. Previous work demonstrated that the properties of the SCs formed in the simulation are in good agreement with observations.  In this paper,  we create mock spectral energy distributions and broad-band photometric images using the radiative transfer code \textsc{skirt 9}.  We test several observational star formation rate (SFR) tracers and find that \mbox{24\,$\mu$m}, total infrared and H$\alpha$ trace the underlying SFR during the (post)starburst phase, while UV tracers yield a more accurate picture of star formation during quiescent phases prior to and after the merger. We then place the simulated galaxy at distances of 10 and 50 Mpc and use aperture photometry at \textit{Hubble Space Telescope} resolution to analyse the simulated SC population.  During the  starburst phase, a hierarchically forming set of SCs leads inaccurate source separation because of crowding. This results in estimated SC mass function slopes that are up to $\sim0.3$ shallower than the true slope of $\sim-1.9$ to $-2$ found for the bound clusters identified from the particle data in the simulation. The masses of the largest clusters are overestimated by a factor of up to 2.9 due to unresolved clusters within the apertures.  The aperture-based analysis  also produces a  relation between cluster formation efficiency and SFR surface density that is slightly flatter than that recovered from bound clusters. The differences are strongest in quiescent SF environments.  

\end{abstract}




\section{Introduction}

Star clusters (SCs) form hierarchically within star-forming interstellar gas clouds and filaments \citep{2003ARA&A..41...57L, 2006ApJ...644..879E, 2017ApJ...840..113G}. The majority of young star clusters have low masses \citep{2019MNRAS.484.4897C, 2020ApJ...893..135M} and form unbound \citep{2021MNRAS.508.5935B}. During their evolution, clusters loose mass and may even dissolve due to the complex interplay of violent relaxation after gas expulsion (\citealt{2006MNRAS.373..752G, 2007MNRAS.380.1589B, 2018A&A...615A..37B, 2020ApJ...900L...4P}), internal dynamics \citep{2008ApJ...679.1272M, 2016MNRAS.458.1450W, 2017MNRAS.469..683T} and external tidal forces in the surrounding galactic environment \citep{1958ApJ...127...17S, 1988ApJ...335..720A, 2011MNRAS.413.2509G, 2013MNRAS.430..676B}. Cluster disruption is strongest at early stages, and young clusters are observed to disrupt at as high a rate as 90\% in number of clusters per a dex in age, as evidenced by the declining age distributions of young clusters in a variety of environments \citep{2010ApJ...719..966C, 2017ApJ...843...91L, 2019MNRAS.484.4897C, 2020ApJ...889..154W}. In later stages, the evolution is more gradual. Recent observations of the LEGUS survey even indicate no evolution in the mass-size relation of clusters older than 100 Myr and that they are not tidally limited nor expanding \citep{2021MNRAS.508.5935B}.

Higher mass clusters that form predominantly bound can therefore remain intact for extended periods of time \citep{2001ApJ...561..751F}.
Such massive SCs in the local universe  have been observed to form in the most complex and extreme star formation environments \citep{1997ApJ...480..235E} such as starbursts and galaxies with disturbed morphologies (e.g. \citealt{1992AJ....103..691H, 1995ApJ...446L...1O, 2018ApJ...869..126L}), interacting galaxies \citep{2003A&A...397..473B, 2010AJ....140...75W} and even dwarf galaxies \citep{1994ApJ...433...65O, 2000AJ....120.1273J, 2011MNRAS.417.1904A, 2021ApJ...912...89K, 2021MNRAS.504.6179E}. The most massive clusters have been suggested as analogues to the progenitors of globular clusters (GCs, see e.g. reviews by \citealt{2010ARA&A..48..431P} and \citealt{2014prpl.conf..291L}). The connection of the present-day young massive clusters to young GCs that typically formed more than 10 Gyr ago in turn requires observations of the sites of GC formation at higher redshifts. Candidates of such regions have been recently observed with ultra deep imaging of fields that contain lensed star-forming clumps beyond $z\sim2$--$3$ \citep{2019MNRAS.483.3618V, 2021A&A...646A..57V}. Significant improvements in spatial resolution in this regard will be provided by the \textit{James Webb Space Telescope} (\textit{JWST}) and the \textit{Extremely Large Telescope}.

The most massive young SCs are often dubbed super SCs (SSCs; \citealt{1971A&A....12..474V, 1985AJ.....90.1163A}) and are typically defined as having masses of or in excess of today's typical GCs ($\gtrsim 10^5$ M$_\odot$, \citealt{2010ARA&A..48..431P}).
The quest for finding these relatively rare objects is, however, complicated by their birth environments. When young, the clusters are surrounded by the obscuring interstellar medium and other nearby clusters \citep{2017ApJ...840..113G} that in projection complicate the determination of the emission of each individual object. Extinction in the starburst environment affects observations in UV-optical wavelengths that otherwise provide the best spatial resolution e.g. with \textit{Hubble Space Telescope} (\textit{HST}). In addition, local (< a few Mpc) starburst galaxies that form such massive clusters (see e.g. \citealt{2006MNRAS.370..513S, 2018ApJ...869..126L}) are rare and their observations are limited by inclination effects and extinction. This has encouraged studies of SSC formation toward larger distances where the larger volume provides both more numerous and more extreme starbursts to be probed.

The compromise of observing outside of the local volume is then that clusters cannot be resolved into individual stars even with the highest resolution instruments such as \textit{HST}. The properties of SCs are instead extracted using fluxes that have been integrated within apertures that cover the bulk of the cluster emission. To counter confusion from nearby objects, observational surveys that rely on aperture photometry use as small apertures as possible and aim at building their SC catalogues with emphasis on single, symmetric and, for example, uniformly colored cluster candidates \citep{2017ApJ...841..131A, 2019MNRAS.484.4897C}. Crowding in the most intense star-forming regions is a major issue and the contamination from the stellar background and nearby objects is minimized by a local sky subtraction. The typical procedure is to estimate the local background using an annulus around the aperture. The small apertures also cause outer parts of the cluster light profile to be truncated at the aperture radius, and the sky subtraction may remove parts of the outer wings of the light profile from the aperture integrated emission \citep{2017ApJ...841..131A}. The underestimated cluster emission has then to be compensated with a filter-specific or cluster-by-cluster estimated aperture correction. In \textit{HST} SC surveys this correction is typically in the range of a fraction of a magnitude \citep{2009ApJS..180...54J, 2010AJ....140...75W, 2017ApJ...841..131A, 2010ApJ...719..966C, 2019MNRAS.484.4897C}, corresponding to tens of percents in the final cluster luminosity or mass estimate. If done using a filter-averaged value for the aperture correction, it is assumed that all clusters have light profiles described by the average over the control sample. The true masses and mass functions of the putative $10^6$--$10^8$ M$_\odot$ SSCs found in recent starburst studies \citep{2017ApJ...843...91L, 2019A&A...628A..60F,  2019MNRAS.482.2530R, 2020MNRAS.499.3267A} are therefore uncertain \citep{2013MNRAS.431..554R}.

Additional challenges are provided by the foreground and background objects, such as point-like stars smeared by the point spread function (PSF), and extended galaxies, as well as single very bright stars in the observed galaxy itself. One of the main observables used to narrow down to the true cluster population  e.g. in \textit{HST} surveys is the concentration index (CI, \citealt{1999AJ....118.1551W, 2017ApJ...841..131A}), which quantifies the concentration of flux, e.g. as a magnitude difference between two concentric apertures with radii of one and three pixels \citep{2010AJ....140...75W, 2017ApJ...840..113G}. Stars can be excluded from the cluster catalogue as having very centrally concentrated flux (low CI), as long as the spatial resolution allows stellar-like and cluster-like light profiles to be separated, while background galaxies are characterized by extended emission (high CI). Recently, \citet{2022MNRAS.509.4094T} introduced an updated formulation for the concentration index which better accounts for the differences in the radial light profiles between PSF-sized stars and extended SCs.

The classification of bona fide clusters out of the catalogues of cluster candidates can then be either computer or human generated, or a combination of both, at varying level of agreement between the two \citep{2021MNRAS.506.5294W}. Isolated, symmetric and separable objects may be extremely challenging to recover in a reliable manner in the 2D projections of intensely star-forming regions and often the central star-forming knots and clumpy cluster candidates are left out of the analysis all together \citep{2002AJ....124.1393L, 2020MNRAS.499.3267A}. 

The challenges outlined above make it difficult to discern the formation process at the high mass end of the cluster mass function (CMF). The main characteristics of massive SCs under discussion are the shape and extent of the the CMF, not to mention the internal properties such as the chemical composition. The search continues for the maximum mass a cluster can reach and if the maximum mass varies with the star formation environment \citep{2006A&A...450..129G, 2008MNRAS.390..759B, 2017ApJ...839...78J}, or if indeed there even exists such a maximum mass \citep{2019ApJ...872...93M}. Galaxies that form stars at a higher rate tend to host brighter SCs \citep{2004MNRAS.350.1503W, 2008MNRAS.390..759B, 2015MNRAS.452..246A} but finding a tight relation is complicated due to low number statistics at the high mass end of the steep (typical power-law slopes of $-2$) mass function of observed SCs \citep{2020MNRAS.499.3267A}. The nature of the cut-off mass is tied to the shape of the CMF as well, as an upper mass limit cut-off would better describe a Schechter-type function rather than a simple power-law \citep{2009A&A...494..539L}. Furthermore, early cluster evolution may affect both the total mass and internal structure of the clusters, and consequently the CMF \citep{2005ApJ...631L.133F, 2006MNRAS.373..752G}. Thus it would be crucial to resolve the SSCs during or immediately after formation, which would for example give a firm base to numerical studies that aim to understand the connection between SSCs and GCs.

Here we take a numerical approach for the detection and analysis of SCs and their environments in a simulated starburst environment.  In previous studies we have analysed in detail the physical properties of the SC population formed in a starburst caused by the merger of two gas-rich dwarf galaxies \citep{2020ApJ...891....2L}. The simulations are a part of the Galaxy Realizations Including Feedback From INdividual massive stars (\textsc{Griffin}) project\footnote{\url{https://wwwmpa.mpa-garching.mpg.de/~naab/griffin-project}} which addresses some current challenges of galaxy formation \citep[see e.g.][for a review]{2017ARA&A..55...59N} allowing for the numerical representation of a resolved ISM \citep[see also][]{2020MNRAS.495.1035S,2022MNRAS.509.5938H}. In the \citet{2020ApJ...891....2L} simulations we have investigated the formation of a distribution of SCs in the filamentary, hierarchical structures of the starbursting merging disks. The high-resolution simulations enable us to take a look at the cloud-scale decoupling of the young stars and SCs from their gaseous environments. The high pressure merger environment together with low shear provided by the low mass of the galaxies enables the formation of hundreds of SCs with masses up to $\sim 10^6$ M$_\odot$. The analysis of the population of bound clusters revealed a CMF that builds with a power-law slope of the order of $-2$ already during the first pericentric passage of the galaxies. Further experiments have shown that the CMF slope is only weakly affected by the star formation efficiency \citep{2022MNRAS.509.5938H}. The population of bound clusters grows in mass and number especially during the most intense star formation period, during which the efficiency of stars forming in bound clusters (cluster formation efficiency, CFE or $\Gamma$, \citealt{2008MNRAS.390..759B}) peaks at $90\%$ of the total star formation. Similar intense cluster formation has recently been reported for example in the HiPEEC\footnote{Hubble imaging Probe of Extreme Environments and Clusters} survey of interacting galaxies of \citet{2020MNRAS.499.3267A}.

Our earlier studies however intentionally utilized the full particle information in the analysis which does not take into account the restricted nature of real observations. Here we aim to process the simulation output to match the data typically used in observational studies of young SCs, and analyse the results using observationally verified methods. In Section 2 we first briefly introduce the simulation setup that has been extensively discussed in \citet{2019ApJ...879L..18L}, \citet{2020ApJ...891....2L} and \citet{2020ApJ...904...71L}. In Section 3 we describe the post-processing and data reduction methods, including the radiative transfer modelling, synthetic photometry and SC detection. In Section 4 we review the spectral energy distribution and extract the star formation properties of the post-processed dwarf starburst. Section 5 investigates the SC population and the brightest objects in the sample, as well as the global cluster formation efficiency, extracted using the \textit{HST}-equivalent photometric detection pipeline. The photometric results are also compared with the underlying bound cluster population. Concluding remarks are given in Section 6.

\section{Simulations}

The simulation analysed here follows an idealised major merger of two gas-rich, low-metallicity dwarf galaxies. The initial conditions, simulation setup and the numerical code have been described in detail in \citet{2014MNRAS.443.1173H, 2016MNRAS.458.3528H, 2017MNRAS.471.2151H} and \citet{2020ApJ...891....2L,2020ApJ...904...71L}. Here we present a brief description of the main aspects of the numerical setup.

We use the \textsc{sphgal} implementation \citep{2014MNRAS.443.1173H} of the widely used \textsc{gadget-3} code \citep{2005MNRAS.364.1105S}. The smoothed particle hydrodynamics (SPH) implementation uses the pressure-energy formulation and the Wendland C$^4$ kernel over 100 neighbours. Artificial viscosity and artificial conduction of thermal energy are included as described in \citet{2013MNRAS.434.3142A}. Our particle mass resolution is approximately $4$ M$_\odot$ per baryonic particle with a $0.1$ pc gravitational softening length. 

We use a chemical network to follow the cooling of gas down to a minimum temperature of $T=10$ K. The network models the abundances of six chemical species (H$_2$, H$^+$, H, CO, C$^+$, O) and free electrons using the reaction rates of H$_2$, H$^+$ and CO as detailed in \citet{2016MNRAS.458.3528H}. Gas above a temperature of $T>3\times 10^4$ K cools according to the metallicity dependent cooling rates of \citet{2009MNRAS.393...99W} for which we follow the mass of H, He, C, N, O, Ne, Mg, Si, S, Ca, Fe and Zn that evolve according to stellar feedback. 

Star formation is modelled using a Jeans mass dependent threshold. Whenever the local Jeans mass of a gas particle in a converging flow crosses eight times the SPH kernel mass ($\sim8\times 400$ M$_\odot$) we allow star formation at a 2$\%$ efficiency per free-fall time. Additionally, in regions where the Jeans mass is resolved with less than half a kernel mass, we enforce instantaneous star formation of any gas particle that exceeds this threshold. \citet{2022MNRAS.509.5938H} has further investigated the effect of varying the star formation efficiency on the cluster formation process. SCs were found to form increasingly less bound  with increasingly higher star formation efficiency. Such loosely bound clusters that were better able to capture the cluster disruption process observed in many galaxies do not, however, match other observed properties of similar mass SCs, such as the half-mass radius or the mean surface density.

The newly formed stellar particles that are used to track feedback processes are sampled into individual stellar masses from the Kroupa IMF \citep{2001MNRAS.322..231K} in the mass range of 0.08 and 50 M$_\odot$. The stellar masses are tabulated in the particle data when the mass drawn is more than one solar mass, in order to trace individually the stellar population that gives the major contribution to the interstellar radiation field (ISRF) discussed below. The IMF sampling is continued as long as the combined mass in the new stellar particle exceeds the progenitor gas particle mass. In the case of stars more massive than $\sim4$ M$_\odot$ we practically model them as individual stellar particles.

The individually tracked stars are then coupled to the surrounding interstellar medium through various radiative and chemical feedback processes. Firstly, all individually stored $>1$ M$_\odot$ stars contribute to the spatially and temporally evolving ISRF through photoelectric heating. Far-ultraviolet (FUV) emission of a given star is integrated from the BaSeL library \citep{2002A&A...381..524W} between 6 and 13.6 eV and propagated to nearby gas particles, taking into account dust and gas column density along the line-of-sight. The background radiation field is set to be the cosmic UV background from \citet{2001cghr.confE..64H}. The young massive stars ($>8$ M$_\odot$) also release photoionizing radiation, producing HII regions modelled as Str\"omgren spheres where overlapping regions are handled iteratively. 
As massive stars reach the end of their lifetime \citep{2013A&A...558A.103G} we model their Type II supernova explosions by injecting the canonical $10^{51}$ erg of thermal energy and the mass and metallicity dependent supernova yield \citep{2004ApJ...608..405C} into the surrounding SPH kernel. Finally, the stars also release asymptotic giant branch winds at a gradual rate according to yields from \citet{2010MNRAS.403.1413K}.

The initial condition for each dwarf galaxy with a virial mass of \mbox{$2\times 10^{10}$ M$_\odot$} includes a dark matter halo with a Hernquist density profile \citep{1990ApJ...356..359H} and $\sim 10^4$ M$_\odot$ mass resolution, and a baryonic disk that constitutes $0.3\%$ of the virial mass, with a $66\%$ gas mass fraction. The gaseous and the stellar disks have masses of $4\times 10^7$ M$_\odot$ and $2\times 10^7$ M$_\odot$ and scale radii of 1.46 kpc and 0.73 kpc.
The parabolic merger setup consists of two identical dwarf galaxies set at an inclined orbit described by initial and pericentric separation of 5 kpc and 1.43 kpc, and inclination and argument of pericentre angles of $i=(60^{\circ}, 60^{\circ})$ and $\omega=(30^{\circ}, 60^{\circ})$.  The initial gas phase metallicity of the simulation is set as $0.1\,\mathrm{Z}_\odot$ and evolves to approximately twice of the initial value by the end of the time span analysed here ($t_\mathrm{sim}=270$ Myr).

\section{Post-processing}

\subsection{Dusty radiative transfer with \textsc{skirt 9}}

We post-process our simulation snapshots with the Monte Carlo radiative transfer code \textsc{skirt 9} \citep{2020A&C....3100381C}. We build the source list based on all stellar particles formed during the simulation. As we track the individual stellar masses down to $1\, \mathrm{M}_\odot$ ($40$--$50\%$ of the total stellar mass) specifically for the purposes of the ISRF within the simulation, we can assign each individual star that contributes to the UV-visual parts of the spectrum with a black body spectrum. The spectra are characterized by the initial mass and age dependent temperature that we extract from the Geneva stellar tracks at corresponding metallicity \citep{2013A&A...558A.103G}. We briefly compare in Section \ref{section:seds} to the population spectra of simple stellar populations (SSP) with respective properties from \citet{2003MNRAS.344.1000B} which are characterized by the initial mass, metallicity and age of the stellar particles.  Atmospheric models with detailed absorption and emission line properties would of course be the most accurate, however based on the qualitative agreement between the SSP and the black body results in our system, we settle with the simplified black body emission. The black body spectra  provide a good approximation of the individually resolved young stellar population dominated by young massive stars.

The dusty interstellar medium is modelled as an octree grid that is built based on the gas density in the simulation. The gas particles are given to \textsc{skirt} as a list of positions, masses, metallicities, SPH-smoothing lengths and temperatures. The dust-tree is allowed to refine down to a scale of $1.5$ parsecs, which is smaller than the typical SPH-smoothing lengths of $\sim2.5$ pc in the cold gas. The gas mass in each SPH-particle is converted into dust mass using the metallicity and a constant dust-to-metals mass fraction of $\sim0.83$, which roughly corresponds to the dust-to-gas mass ratio of $0.1\%$ used in the simple radiative transfer performed during the simulation. For the dust composition and emission properties, we assume the Milky Way dust distribution from \citet{2001ApJ...548..296W} and \citet{2001ApJ...554..778L} which includes a mixture of polycyclic hydrocarbon, silicate and graphite dust grains. The radiative transfer in \textsc{skirt 9} could additionally take into account the self-absorption of the dust emission iteratively. We tested its effect on the most intense star formation phases, where the resulting self-absorption was at a level of a few percent, and leave it out of our analysis as our dust densities are still relatively low.

\textsc{skirt 9} includes also built-in tools to transport the output emission to a cosmological distance. For the majority of the analysis we assume a redshift of zero and only position the merger at a relevant distance of a few tens of megaparsecs. In Section \ref{section:seds} we briefly discuss the merger in relation to the sensitivity of the \textit{JWST}, where we  use the the Planck 2018 cosmology of $\Omega_m=0.315$ and $h=0.674$ \citep{2020A&A...641A...6P} to move the starburst to redshift $z=0.1, 0.5, 1$ and $1.5$. The output is a redshifted spectrum with cosmological surface brightness dimming applied. Heating by the cosmic microwave background could also be applied to these synthetic observations, however at the redshifts considered here its contribution to the gas temperature is negligible and we exclude it from our post-processing.

\subsection{Synthetic images} \label{section:filtering}

The most relevant observational counterparts to our present study span from UV to IR. We make use of the instrument throughputs included in \textsc{skirt 9} by producing images in the following filters: the standard Johnson-Cousins UBVRI, the \textit{Spitzer Space Telescope} infrared broad-band filters at $24\,\mu$m, $70\,\mu$m and $160\,\mu$m, and rest-frame FUV and NUV filters of the \textit{Galaxy Evolution Explorer} (\textit{GALEX}). The standard UBVRI filtered images also represent here the corresponding \textit{HST} observations as the images taken with the various versions of the slightly different visual \textit{HST} filters are often used interchangeably with the UBVRI system \citep{1999AJ....118.1551W,  2010AJ....140...75W, 2020MNRAS.499.3267A}. We also use the \textsc{python} package \textsc{pyphot} to extract \textit{JWST} NIRCam images in the redshifted UV/visual, keeping in mind the wavelength and sensitivity range of the NIRCam filters\footnote{https://jwst-docs.stsci.edu/near-infrared-camera/nircam-predicted-performance/nircam-sensitivity}. When discussing star formation tracers, we use the combined $24\,\mu$m, $70\,\mu$m and $160\,\mu$m fluxes to approximate for the total infrared (TIR) luminosity following \citet{2002ApJ...576..159D} as 
\begin{equation}
    L_\mathrm{TIR} = 1.559\nu L_\nu(24\,\mu\mathrm{m}) + 0.7686\nu L_\nu(70\,\mu\mathrm{m}) + 1.347\nu L_\nu(160\,\mu\mathrm{m}) .
\end{equation}
  
\begin{figure*}
	\includegraphics[width=0.9\textwidth]{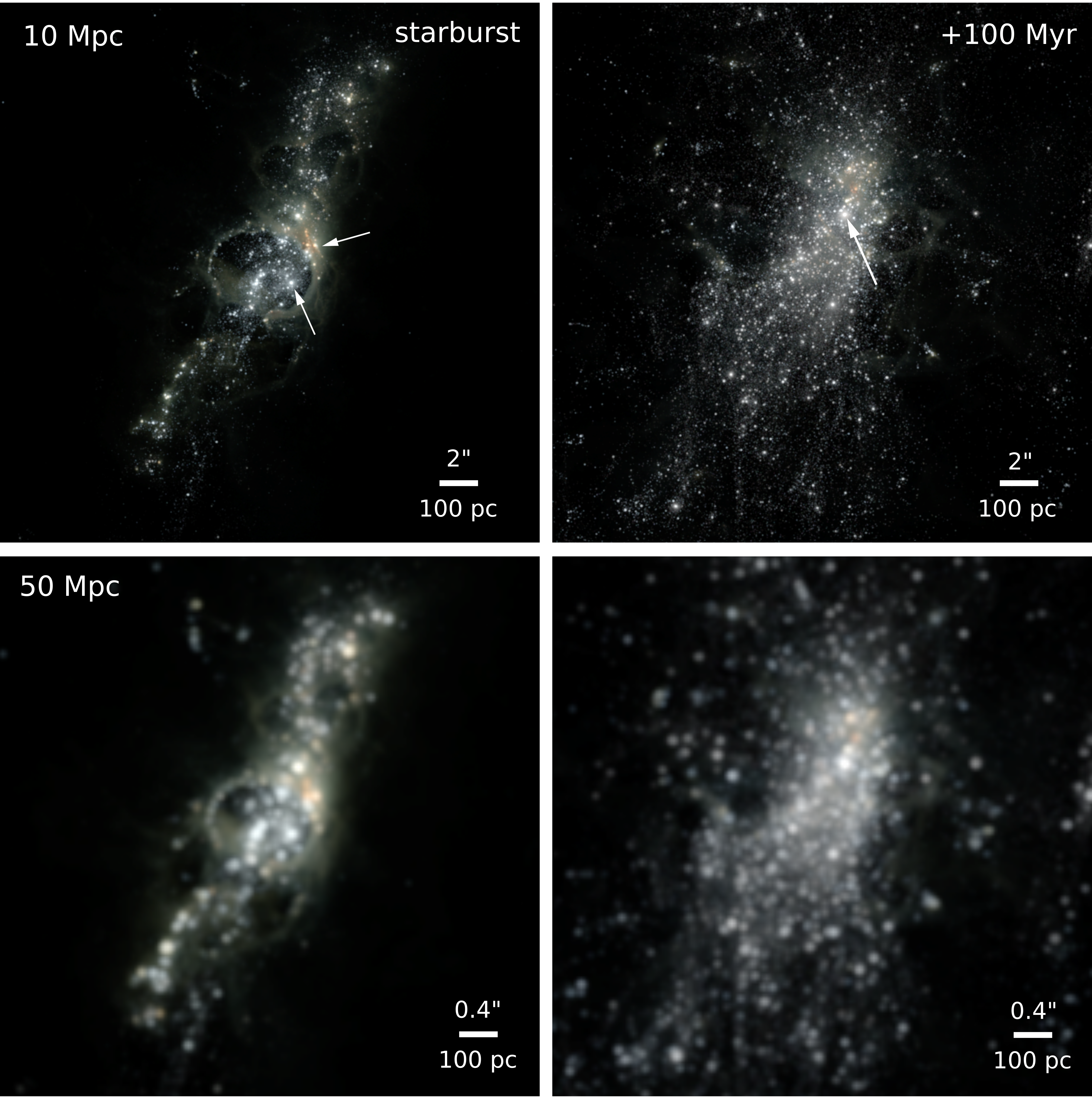}
    \caption{Color composite images in B (blue), V (green) and I (red) bands of the starburst (left, t$_\mathrm{sim}=169$ Myr) and 100 Myr later (right) at a distance of 10 Mpc (top) and \mbox{50 Mpc} (bottom). The images span 1.4 kpc in side, the original image resolutions are 1.5 (top) and 7.5 (bottom) pc per pixel, and each band has been smoothed with a Gaussian PSF of FWHM=2.1 pixels that is a combination of the \textit{HST} image resolution of 0.04$\arcsec$ per pixel and a typical PSF of FWHM=1.6 pixels. The fluxes have a logarithmic stretch. For clarity, no noise has been added to these images. In the left panels, the gaseous and reddened shell wall (top arrow) of the superbubble is produced by the clusters in the couple hundred pc wide cavity (bottom arrow). The obscured region hosts the most intense starburst where the most massive SC in the entire simulation is in the process of building its stellar mass, hierarchically, through the formation and coalescence of tens of smaller mass SCs. The most massive cluster is consequently visible in the right hand panels as the brightest cluster (arrow), surrounded by the very irregular distribution of lower mass clusters. The images are best viewed on a computer screen.}
    \label{fig:BVI}
    \centering
\end{figure*}

The typical SC surveys are done with e.g. \textit{HST} ACS/WFC and WFC3 UVIS channels at a resolution of $0.04$--$0.05\arcsec$ and typical point spread functions (PSF) of the order of 0.06--0.1 $\arcsec$ \footnote{e.g. https://hst-docs.stsci.edu/wfc3ihb/chapter-6-uvis-imaging-with-wfc3/6-6-uvis-optical-performance}, and with the upcoming \textit{JWST} NIRCam at $0.031$ -- $0.063\arcsec$ and PSFs of 0.03--0.16 $\arcsec$ \footnote{https://jwst-docs.stsci.edu/jwst-near-infrared-camera/nircam-predicted-performance/nircam-point-spread-functions}. The spatial resolutions are therefore quite similar and we simplify our analysis by using a default pixel scale of $0.031\arcsec$ that we can scale to lower resolution when applying a PSF. We position the merger at two fiducial distances of $10$ and $50$ Mpc (see Section \ref{section:ap_phot}), which correspond to pixel scales of $1.5$ and $7.5$ pc. We produce broadband images of the full spectral energy datacubes along the merger through the various filters using the built-in instruments of \textsc{skirt}. Observational effects such as noise and PSF are added when relevant according to the instrument and explained in more detail in the corresponding analysis section. Example color composite images composed of the filtered \textsc{skirt} output are shown in Fig. \ref{fig:BVI} at the time of the most intense starburst and 100 Myr after the starburst.

We also approximate the spatial H$\alpha$ flux by extracting the H$\alpha$ emission from recombination and collisional excitation of hydrogen in HII regions. For this we follow \citet{2013ApJ...779....8K} and \citet{2017MNRAS.466.3293P} and calculate the three dimensional emissivity on a grid using the formulation for recombination from \citet{2011ApJ...727...35D} as given by 
\begin{equation}
    dL_{\mathrm{H}\alpha, R} = 4\pi \times 2.82 \times 10^{-26}\, T_4^{-0.942-0.031\ln{T_4}}\, n_\mathrm{e} n_\mathrm{H+} dV
\end{equation}
from volume element $dV$ in units of erg s$^{-1}$. Here $T_4$ is the gas temperature $T$ in units of $10^4$ K, and $n_\mathrm{e}$ and $n_\mathrm{H+}$ are the electron and ionized hydrogen number densities. For collisional de-excitation from $n=3\rightarrow2$ we follow \citet{2013ApJ...779....8K} and use 
\begin{equation}
    dL_{\mathrm{H}\alpha, C} = 1.30\times 10^{-17} \frac{\Gamma_{13}(T)}{\sqrt{T}} \exp \left(\frac{-12.1\mathrm{eV}}{k_\mathrm{B} T} \right) n_\mathrm{e} n_\mathrm{H} dV
\end{equation}
where $k_\mathrm{B}$ is the Boltzmann constant, $n_\mathrm{H}$ is the number density of neutral hydrogen and $\Gamma_{13}$ is the effective collision strength from $n=1\rightarrow 3$ given by a polynomial expression
\begin{equation}
    \Gamma_{13}(T) = 0.35 -2.62 \times 10^{-7} T - 8.15\times 10^{-11} T^2 + 6.19\times 10^{-15} T^3  
\end{equation}
between $4\times 10^3$ K and $2.5\times 10^4$ K and 
\begin{equation}
    \Gamma_{13}(T) = 0.276 +4.99 \times 10^{-6} T - 8.85\times 10^{-12} T^2 + 7.18\times 10^{-18} T^3  
\end{equation}
between $2.5\times 10^4$ K and $5\times 10^5$ K according to \citet{1983MNRAS.202P..15A}. We exclude the contribution from the diffuse interstellar gas by ignoring gas below a number density of $10$ cm$^{-3}$ or above a temperature of \mbox{$10^{5.5}$ K}.

The contributions from recombination and collisional de-excitation are then projected onto a map with similar extension and resolution as used in the broadband photometry to enable H$\alpha$ photometry. As the H$\alpha$ emission in observational studies is often corrected for extinction, we take the unattenuated H$\alpha$ emission as our SFR proxy and compare this to the other dust-affected and corrected SFR tracers as explained in the next Section.

\subsection{Star formation rate tracers}

The processed spectra provide us with observational estimates of the star formation rates in the merger, which can be compared to the values obtained directly from the snapshots. We use the broadband IR fluxes and TIR estimates in combination with the UV fluxes to produce an attenuation corrected SFR estimate following \citet{2011ApJ...741..124H}. First, for the SFR estimates in the various bands we use the general form 
\begin{equation}\label{eq:sfr_tracer}
    \mathrm{SFR}(\mathrm{band})\, [ M_\odot\, \mathrm{yr}^{-1}] = C_1 \times L(\mathrm{band})
\end{equation}
where $L=\nu L_\nu$ from broad-band photometry is in units of erg s$^{-1}$, and the calibration constants are listed in Table \ref{tab:sfr_calibration}. 
For the attenuation corrected FUV and NUV based SFR estimates we use a form
\begin{equation}
    L(\mathrm{UV})_\mathrm{corr}= L(\mathrm{UV})_\mathrm{obs} + C_2 L(\mathrm{IR})_\mathrm{obs}
\end{equation}
with the constants listed again in Table \ref{tab:sfr_calibration}. The $24\,\mu$m correction was originally calibrated for the \textit{InfraRed Astronomical Satellite} (\textit{IRAS}) $25\,\mu$m filter that has been shown to be interchangeable with the \textit{Spitzer} $24\,\mu$m filter we use here \citep{Kennicutt_2009}. As we do not include the very old stellar population or the very low mass stars ($<1$ M$_\odot$), we also do not need to correct the IR emission for emission that is not specific to star formation.

\begin{table}
	\centering
	\caption{The calibration factors and flux corrections used when converting from broadband and H$\alpha$ luminosity to SFR.}
	\label{tab:sfr_calibration}
	\begin{tabular}{lcc}
		\hline
		Band & $\log{C_1}$ & Reference\\
		\hline
		FUV & -43.35 & \citet{2011ApJ...741..124H} \\
		NUV & -43.17 & \citet{2011ApJ...741..124H} \\
		TIR & -43.41 & \citet{2011ApJ...741..124H} \\
		$24\,\mu$m & -42.69 & \citet{2009ApJ...692..556R} \\
		H$\alpha$ & -41.26 & \citet{2010ApJ...714.1256C}, \citet{2011ApJ...741..124H} \\
		\hline
		\end{tabular}
		\begin{tabular}{lcc}
		Band & $C_2$ & L(IR) \\
		\hline
		FUV & 0.46 & TIR \\
		NUV & 0.27 & TIR \\
		FUV & 3.89 & $24\,\mu$m\\
		NUV & 2.26 & $24\,\mu$m \\
		\hline
	\end{tabular}
\end{table}

\subsection{Detection of star clusters and aperture photometry}\label{section:ap_phot}

We connect our simulated dwarf starburst to local star-forming and starburst galaxies where SC formation has been widely studied, such as the LEGUS dwarf sample \citep{2019MNRAS.484.4897C} within \mbox{$12$ Mpc}, the Antennae at \mbox{$20$ Mpc} or the merging galaxies in the HiPEEC survey at \mbox{$\sim 50$ Mpc}. We take an approach where we position the simulated system across its evolution to similar distances, namely \mbox{10 Mpc} and \mbox{50 Mpc}. The stars within SCs cannot be resolved at such distances and the integrated emission extracted with aperture photometry at the location of the clusters is used instead to quantify the properties of the clusters. Following the general ideas of cluster detection in recent observational surveys we need to identify the SCs and extract their  aperture-averaged properties.

We start with the B, V and I band images, that are commonly analysed in photometric cluster catalogues, and smooth over the images with a Gaussian PSF of FHWM$=2.1$ pixels to emulate the spatial resolution of \textit{HST} at $0.04\arcsec$  per pixel combined with a typical PSF of FWHM=1.6 pixels. We add a noise field of $32$ nJy per pixel ($\sim27.6$ ABmag, or $\sim27$ ABmag translated to \textit{HST} resolution of 0.04 \arcsec per pixel) with a $10\%$ standard deviation. The noise is added to mimic the typical sensitivity limit of 1 hour long exposures with \textit{HST}\footnote{https://hst-docs.stsci.edu/wfc3ihb/chapter-6-uvis-imaging-with-wfc3/6-8-uvis-sensitivity} that are typical for SC surveys. The background estimation is performed in patches of 15 pixels using the SourceExtractor methodology of \textsc{Background2D} in the \textsc{photutils} tools in \textsc{astropy}. The cluster catalogue is then built by identifying continuous structures in the image with at least three pixels $5\sigma$ above the artificial background using the detection and deblending routines in \textsc{photutils}. We tested a few contrast parameter values between $0.01$-$0.0001$ to allow easy separation of brightness peaks in crowded regions but did not find the results to be sensitive to the exact value when going beyond the standard value of $0.001$.

In the two fiducial resolution images of $1.5$ and $7.5$ pc per pixel, we use aperture size of 3 pixels that correspond to $4.5$ and $22.5$ pc. These are typical aperture sizes used in  nearby SC studies such as the LEGUS dwarf survey and studies of the Antennae (2--6 pixels, from $\sim$ a few pc to $15$ pc aperture radii, \citealt{2010AJ....140...75W, 2019MNRAS.484.4897C}) and slightly more distant surveys such as the HiPEEC and the GOALS (typically 3 pixels, $> 10$ pc, \citealt{2017ApJ...843...91L, 2020MNRAS.499.3267A}), respectively. The local sky background is removed as the median flux within a one pixel wide annulus from 5 to 6 pixels centered at the aperture location. The aperture correction that accounts for the missing light in the outer wings of each cluster and the light in the wings removed by the sky annulus is estimated using flux growth curves of isolated clusters. In practice the aperture correction is only possible to perform in the 10 Mpc images, as there are not enough suitable isolated clusters to get a statistically meaningful estimate in the 50 Mpc images. The integrated fluxes of all detected clusters in the 10 Mpc images are corrected by the median of the ratio between the flux within 3 and 10 pixels, centered at each isolated control cluster. The correction is typically only $\sim 0.1$ ABmag ($\sim +10 \%$ in flux), reflecting the small size of the majority of the clusters (see e.g. Fig. 9 of \citealt{2020ApJ...891....2L} and discussion in Section \ref{section:candidates}). For the brightest isolated clusters the individual aperture corrections would be as high as 0.5 ABmag ($\sim +60 \%$). However, we adopt the averaged value in each image as the value of the correction as it cannot be estimated individually for all of the clusters.

The dwarf system is fairly small compared to disk galaxies more typically analysed in starburst studies. Crowding can be a problem in the central region, especially during the starburst itself. To prevent double counting from overlapping apertures, we exclude apertures that overlap by more than the aperture radius and only include the brightest aperture in each region where overlapping occurs. This is similar to for example the HiPEEC study, where they exclude apertures whose centres are closer than the aperture radius.
This means that in the better resolution image we will have a larger number of apertures (i.e. cluster candidates) both due to resolution (less blending) and aperture size (lower chance of overlapping). Additionally, we follow observational cluster catalogues and require each aperture to be detected in B, V and I bands within two pixels. Finally, we go through the resulting catalogue by hand and remove objects that do not have a concentrated, cluster like flux distribution.

The integration and corrections for each aperture can then be performed both over the filtered photometric images, but also over the actual underlying particle data. Based on earlier work, we know the full star formation properties and cluster populations in each snapshot, thus we can compare the photometrically extracted cluster population to the real population underneath. Briefly, we consider in the bound cluster population those clusters that survive through friend-of-friends and \textsc{sufbind} procedures \citep{2001MNRAS.328..726S, 2009MNRAS.399..497D} with at least 50 bound star particles, i.e. roughly 200 M$_\odot$.

\section{Panchromatic view of a dwarf galaxy starburst}

\subsection{Spectral energy distribution}\label{section:seds}

\begin{figure*}
	\includegraphics[width=0.9\textwidth]{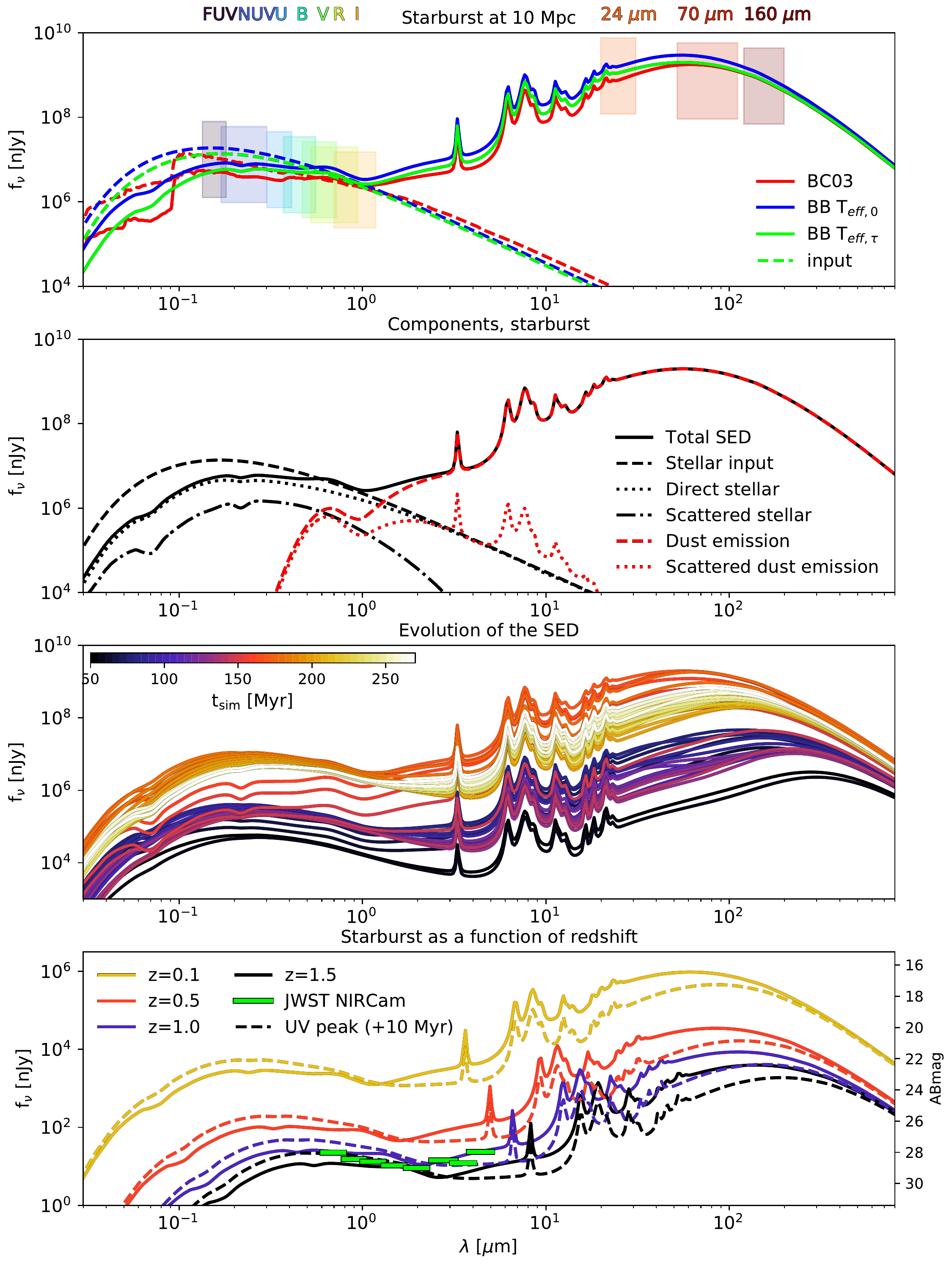}
    \caption{The spectral energy distribution of the merging galaxies. The top panel compares three types of input spectra at the time of the starburst: the stellar population spectra  of \citeauthor{2003MNRAS.344.1000B} (\citeyear{2003MNRAS.344.1000B}, BC03, red), the black body spectra assuming the zero-age properties for all stars ($T_\mathrm{eff,0}$, blue) and the black body spectra following stellar evolution tracks from \citeauthor{2013A&A...558A.103G} (\citeyear{2013A&A...558A.103G}, green) which is the fiducial input model in the rest of the panels. The broadband filters used in the analysis are also indicated. The second panel shows an example of the components that result in the full output SED at the time of the peak starburst. The third panel shows the evolution of the composite spectrum from the first passage \mbox{($\sim 50$ Myr)}, through the starburst ($\sim 150-190$ Myr) and up to 100 Myr past the starburst in steps of 5 Myr. The bottom panel shows the SEDs of the starburst (solid) and the system when the peak in UV-flux is reached (dashed), transferred to redshifts of $z=0.1$ (yellow), $0.5$ (red), $1.0$ (blue) and $1.5$ (black) in comparison with the expected \textit{JWST} NIRCam filter sensitivities for point sources in 10 000 sec exposures (horizontal lime green bars). See text for more details.}
    \label{fig:SEDs}
    \centering
\end{figure*}

In Fig. \ref{fig:SEDs} we investigate the integrated spectral energy distribution (SED) of the simulated merger that evolves according to the (attenuated) emission of the resolved stellar population. The integrated SEDs are produced by summing over the flux in all pixels in an image at each wavelength, separately in each snapshot that are produced every 5 Myr. We concentrate mostly on the spectrum at the time of the most intense starburst. The top panel shows a comparison of the integrated galaxy SED at the time of the peak starburst using either the SSP spectra or the black body spectra as input for the resolved stars. We compare the assumption of a single initial temperature of each stellar track to following the detailed temperature evolution of each individual star when calculating the black body spectra. We show the input spectra with dashed lines and the corresponding final SEDs  as solid lines. The results using SSP spectra \citep{2003MNRAS.344.1000B} as input are shown in red and the black body models that account for stellar evolution \citep{2013A&A...558A.103G} in green and using fixed initial temperature in blue. The top panel also shows the wavelength range of the broad-band filters used in the photometric analysis. 

Overall the resulting fluxes using evolving black bodies and SSPs show quite similar shapes, within a factor of 2. The simplified black body models, however, better capture the spatial distribution of hard emission localised around the individually resolved young massive stars, compared to assigning each stellar mass particle with a SSP spectrum that only specifies population averaged emission. The black body models result in slightly higher integrated flux at UV-visual wavelengths, that then through absorption and re-emission translates into a higher flux of integrated dust emission. A similar qualitative agreement between the SSP and black body models applies in other phases of the merger as well, only at a lower level of emission as discussed below.

\begin{figure}
	\includegraphics[width=\columnwidth]{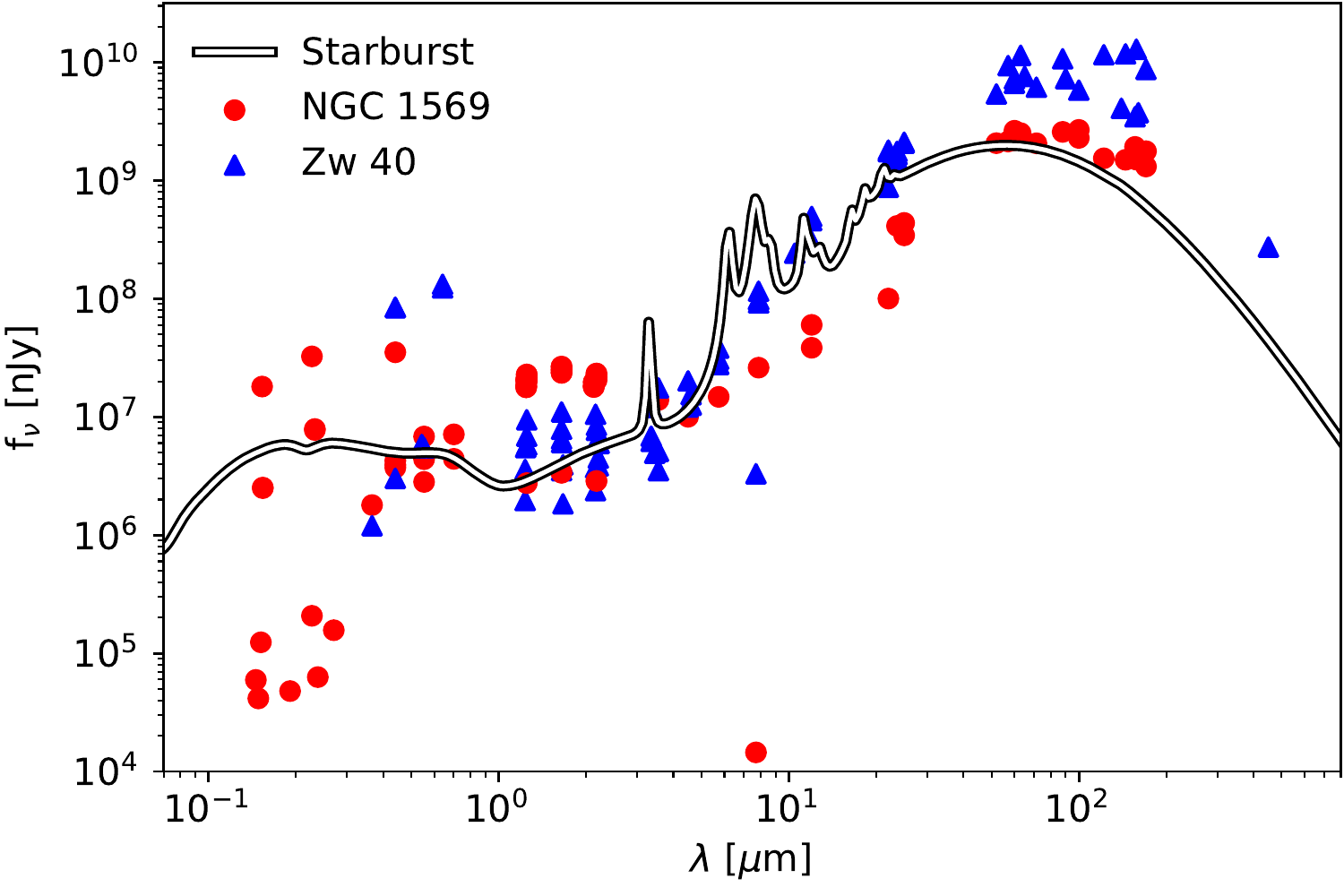}
    \caption{The starburst SED compared with broadband photometric observations of dwarf galaxies NGC 1569 and II Zw 40 that have recently undergone a starburst. The photometric observations show some scatter especially at shorter wavelengths, and our simulated starburst settles among the observed results.}
    \label{fig:SED_dwarfs}
    \centering
\end{figure}

The second panel shows the components that result in the integrated galaxy SED (solid black line) at the time of the starburst, using the evolving black body spectra of the resolved stars as input (dashed black line). The components shown are the direct, dust attenuated stellar emission (dotted black), stellar emission that has been scattered by dust (dot-dashed black), direct emission of dust that has been heated by the stellar spectra (dashed red), and the dust emission that has been scattered by dust (dotted red). At UV to visual wavelengths the spectrum is dominated by attenuated emission from the stars, while at longer wavelengths the dust emission dominates. The stellar and dust emission that are scattered by dust only give minor contributions due to the relatively low gas densities. The spectral features between 1 and 30 micron are characteristic of the different dust species.

The third panel follows the integrated galaxy SED at various epochs across the merger. At early times, closer to the first pericentric passage, gas in the system is quite uniformly distributed, and the dust can remain cool as absorption is low. The peak wavelength of the IR emission, often measured e.g. as the flux ratio between short and long wavelength IR bands (see the Spitzer bands in the top panel), is characterised by the dust temperature (see e.g. \citealt{2018A&A...609A..30S}) and moves up and down following the intensity of star formation (see Section \ref{section:SFR_tracers}). During the starburst (150--190 Myr), the infrared contribution flares up as the majority of the stars and consequently SCs (see Fig. 5 of \citealt{2020ApJ...891....2L}) form, partly obscured by the clouds as shown in Fig. \ref{fig:BVI}. This phase however only lasts a few tens of Myrs. The shape of the SED returns again towards that of the earlier phases after the starburst, when the dominating clustered stellar light has emerged from the birth clouds. Dust temperature also decreases after the starburst, however not as low as before the starburst as the star formation rate of the post-merger system still remains at an order of magnitude higher level.

The bottom panel of Fig. \ref{fig:SEDs} shows the integrated starburst SED (solid) which represents the epoch of maximum infrared emission translated to various redshifts up to $z=1.5$. We also show the SED of the merger when it reaches its peak UV emission (peak of starburst plus 10 Myr, dashed, see next Section for discussion of the UV emission), to capture the maximal emission in the short wavelength \textit{JWST} bands. The horizontal bars indicate the sensitivity of the broad-band NIRCam filters where a 10 000 second exposure is assumed, with the highest sensitivity corresponding to 29 ABmag. 

Assuming a point-source origin for the SED in the bottom panel, we would be able to catch the dwarf starburst roughly out to a redshift of $z\sim1$ and during the UV-peak, at a total rest-frame FUV magnitude of roughly $-16$, in a couple of bands at redshift $z=1.5$. The image resolution at redshift $z\sim1$ is more than 200 pc per pixel. The rest-frame continuum FUV magnitude around the most massive cluster is between $-14$ and $-15$ ABmag depending on the area (radii from $\sim10$ pc to 200 pc). Therefore, after applying a Gaussian band-specific PSF ($0.987$--$2.3$ pixels), none of the individual pixels at $z=1$ exceed the 10 000 second exposure sensitivity. The brightest pixels reach barely 30 ABmag in the \textit{JWST} filters at $z=1$ and 28.8 ABmag at $z=0.5$. Detecting the stellar continuum of this system even in a few \textit{JWST} pixels at redshifts beyond $z\sim0.5$ would require a longer exposure such as the more than one magnitude deeper JADES survey (expected sensitivity down to 2.8 nJy) planned for the \textit{JWST} which will exhibit 20 hour exposures \citep{2020IAUS..352..342B, 2020ApJ...892..125H}. Such deep simulated observations have been recently discussed for example in \citet{2021ApJ...913L..25G}, however from a dwarf galaxy point of view rather than SSC formation. Alternatively, observations of gravitational lensing with \textit{JWST} of such star-forming knots will provide a unique tool to resolve such compact objects at higher redshifts. Proto-GCs have been resolved beyond redshift $z\sim6$ and rest-frame magnitudes beyond 30 ABmag already with \textit{HST} and the Multi Unit Spectroscopic Explorer instrument at the Very Large Telescope \citep{2017MNRAS.467.4304V, 2021A&A...646A..57V} at a resolution of tens of parsecs.

Here we have concentrated on the continuum emission of stars and dust in a merging dwarf galaxy system. Nebular continuum emission and spectral line emission, such as the H$\alpha$ and the [OIII] 5007 \r{A}ngrst\"om lines, may however also contribute significantly in such low mass star-forming galaxies (e.g. \citealt{2016ApJ...819...73L, 2021ApJ...916...11I}). Both of these lines would be observed in the shortest wavelength NIRCam bands at redshifts of $z\sim0.5$--$2$. The increase in flux would however need to be a factor of 2--3 at $z\sim 0.5$ and 6--7 at $z\sim 1$ even in the brightest pixels to make the starburst or its UV-peak emission detectable in multiple \textit{JWST} bands in a 10 000 sec exposure. In Section \ref{section:SFR_tracers} we show that the intrinsic H$\alpha$ contributes only a small fraction to the integrated flux of the system. Combined, all the nebular lines and the nebular continuum could add up to a factor 5 more emission compared to the stellar emission \citep{2017ApJ...840...44B} in the optical and near infrared wavelengths that are not dominated by dust emission and that are redshifted into the \textit{JWST} bands at $z\sim 0.5$--1. Inclusion of the nebular emission could therefore increase the flux in the \textit{JWST} bands and make the starburst region detectable in the most sensitive bands at least to $z\sim 0.5$.

\subsection{Comparison to observed dwarf starbursts}

For the observational perspective, we compare in Fig.  \ref{fig:SED_dwarfs} our starburst SED to archival photometry of two dwarf galaxies, NGC 1569 and II Zw 40, obtained from the The NASA/IPAC Extragalactic Database\footnote{https://ned.ipac.caltech.edu/}. These dwarf galaxies have very similar gas and dust masses to our system and, most importantly, are actively forming stars. The dust properties of these galaxies have been discussed in \citet{2003A&A...407..159G, 2005A&A...434..867G} and the star formation history of NGC 1569 in \citet{2005AJ....129.2203A} and of II Zw 40 in \citet{1996ApJ...466..150V}. II Zw 40 has been undergoing a starburst for the past few Myr with a SFR of $1.2$--$1.4$ M$_\odot\,\mathrm{yr}^{-1}$ caused by a merger \citep{Kepley_2014, 2018ApJ...865...55L}, indicated by tidal tails seen in optical and IR. NGC 1569 has had a few star formation episodes within the past ten Myr at a level of $0.01$--$0.1$ M$_\odot\,\mathrm{yr}^{-1}$. II Zw 40 has a metallicity of $\sim 1/6$--$1/5$ Z$_\odot$, best-fit dust-to-metals and gas-to-dust ratios of $530$--$1460$ and $1/5$--$1/2$, approximately $5$ to $10$ times more HI than our system and very little molecular hydrogen \citep{2001AJ....121..740M, 2016ApJ...828...50K}. NGC 1569, at slightly higher metallicity of $\sim 1/4$ Z$_\odot$, has best-fit dust-to-metals and gas-to-dust ratios of $740$--$1600$ and $1/6$--$1/3$ and a very similar HI mass to ours. Both of these systems include SCs in the SSC mass range ($>10^5$ M$_\odot$, \citealt{2004MNRAS.347...17A, 2008ApJ...686L..79G, 2018ApJ...865...55L}) that have ages that coincide with the bursts of recent star formation. We have translated the fluxes of NGC 1569 from the distance of 2.2 Mpc \citep{1988A&A...194...24I} and of II Zw 40 from 10.5 Mpc \citep{1988cng..book.....T} to our fiducial 10 Mpc for Fig.  \ref{fig:SED_dwarfs}. 

The SED of our merger, that has peak SFR values at a few \mbox{$\sim 0.1$ M$_\odot\mathrm{yr}^{-1}$}, gas-to-dust ratio of $1/0.001=1000$, and somewhat lower metallicity, falls right among the observed photometric results of these two observed systems. The shape of the dust emission in both galaxies coincides with our resulting SED. II Zw 40, which has a higher gas mass and star formation rate, shows higher rates of IR emission at long IR-wavelengths. The scatter of the datapoints recovered from the archive is however considerable, especially at shorter wavelengths where corrections for Milky Way dust can introduce a lot of uncertainty. Note also that we do not include spectral lines in our SED modelling. Strong continuum and H$\alpha$ or [OIII] line emission, for instance, would result in increased flux at sub-micron broad-bands. Overall the simulated results agree well with the photometric properties of observed low-metallicity dwarf starbursts.

\subsection{Inferred star formation rate}\label{section:SFR_tracers}

\begin{figure}
	\includegraphics[width=\columnwidth]{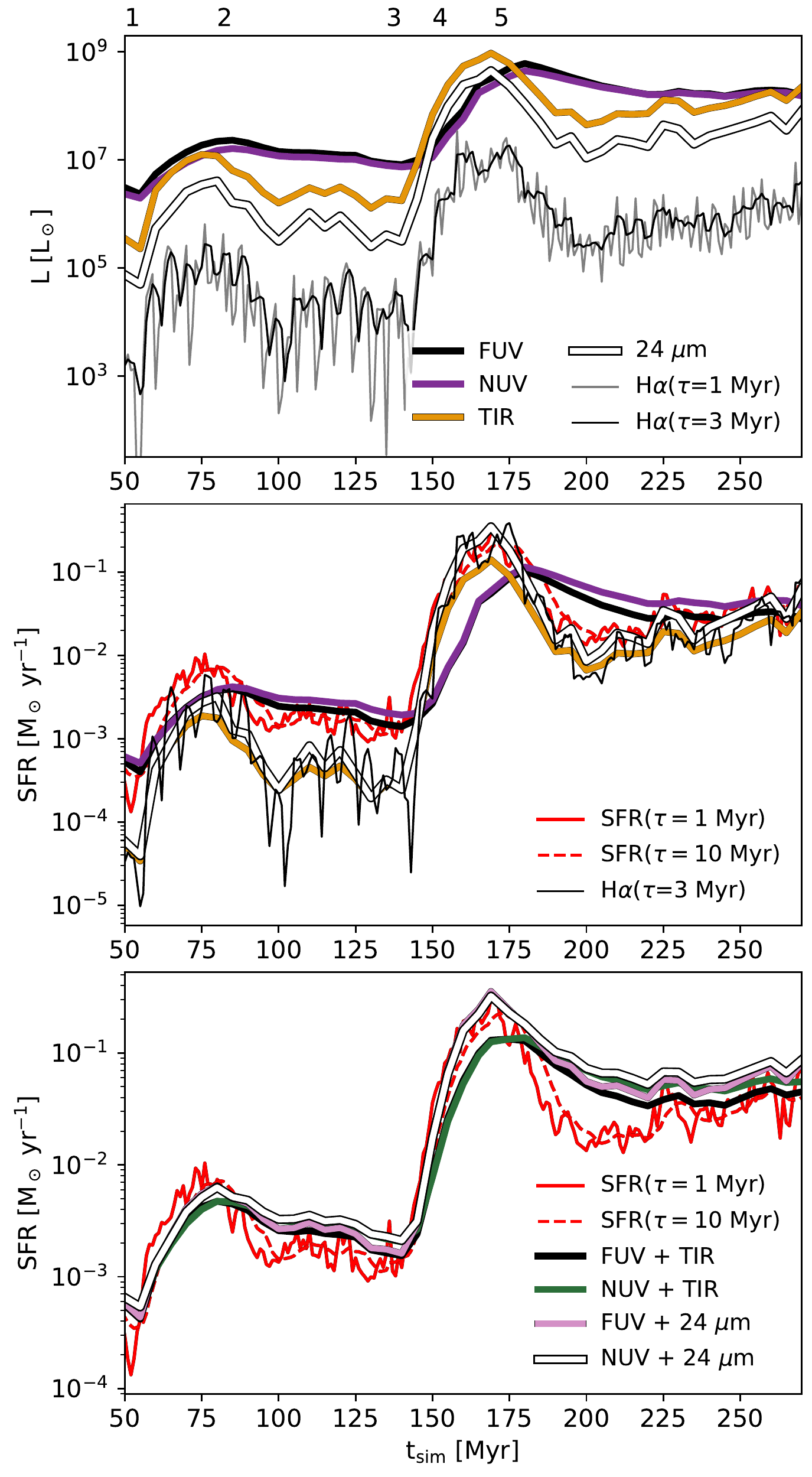}
    \caption{Top: the integrated luminosity during the merger in various dust affected broad bands and from direct H$\alpha$ emission. The integrated broad band emission in FUV (thick black), NUV (purple), TIR (orange) and 24 $\mu$m (white) are shown every 5 Myr and the H$\alpha$ is averaged over the past 1 Myr (gray) and 3 Myr (thin black). Middle: the luminosities translated into single-band star formation rates according to the calibrations given in Table \ref{tab:sfr_calibration}, compared with the star formation rate directly from the stellar data (past 1 Myr average in solid red; past 10 Myr average in dashed red). The H$\alpha$ is only shown as a 3 Myr average that best reflects the mean age of the ionizing stellar population. Bottom: star formation rates from the IR corrected UV luminosities. The numbers at the top through 1 to 5 denote approximately the following phases during the merger: the first passage (1), the first apocentre (2), the second passage and gas disk coalescence (3), the onset of the starburst (4), and the blowout of the central star-forming region (5, see Fig. 2 of \citealt{2020ApJ...891....2L} for visual reference).}
    \label{fig:SFRs}
    \centering
\end{figure}

In Fig. \ref{fig:SFRs} we show the integrated luminosities (top panel) in various bands and the corresponding star formation rate estimates (middle and bottom panels) during the merger. For comparison we also show the SFR extracted directly from the snapshots using either very young stars with ages less than 1 Myr or averaging over a bit longer time span using stars younger than 10 Myr. The luminosities have been translated from the total flux in each photometric/H$\alpha$ map using the calibrations given in Table \ref{tab:sfr_calibration}. Most of the tracers are sensitive to changes in the SFR over the past couple tens of Myr, while only the H$\alpha$ emission is able to respond to changes at a time scale of a few Myrs \citep{2012ARA&A..50..531K} as it is sensitive to only the highest mass stars. We therefore show the time evolution sampled every \mbox{5 Myr} in the broad-band tracers which naturally smooth out variations on short timescales, and show the more time-sensitive H$\alpha$ tracer averaged over past 3 Myr. The average of 3 Myr reflects the mean age of stars that contribute to the ionizing radiation \citep{1999ApJS..123....3L, 2012ARA&A..50..531K}.

\begin{figure*}
	\includegraphics[width=\textwidth]{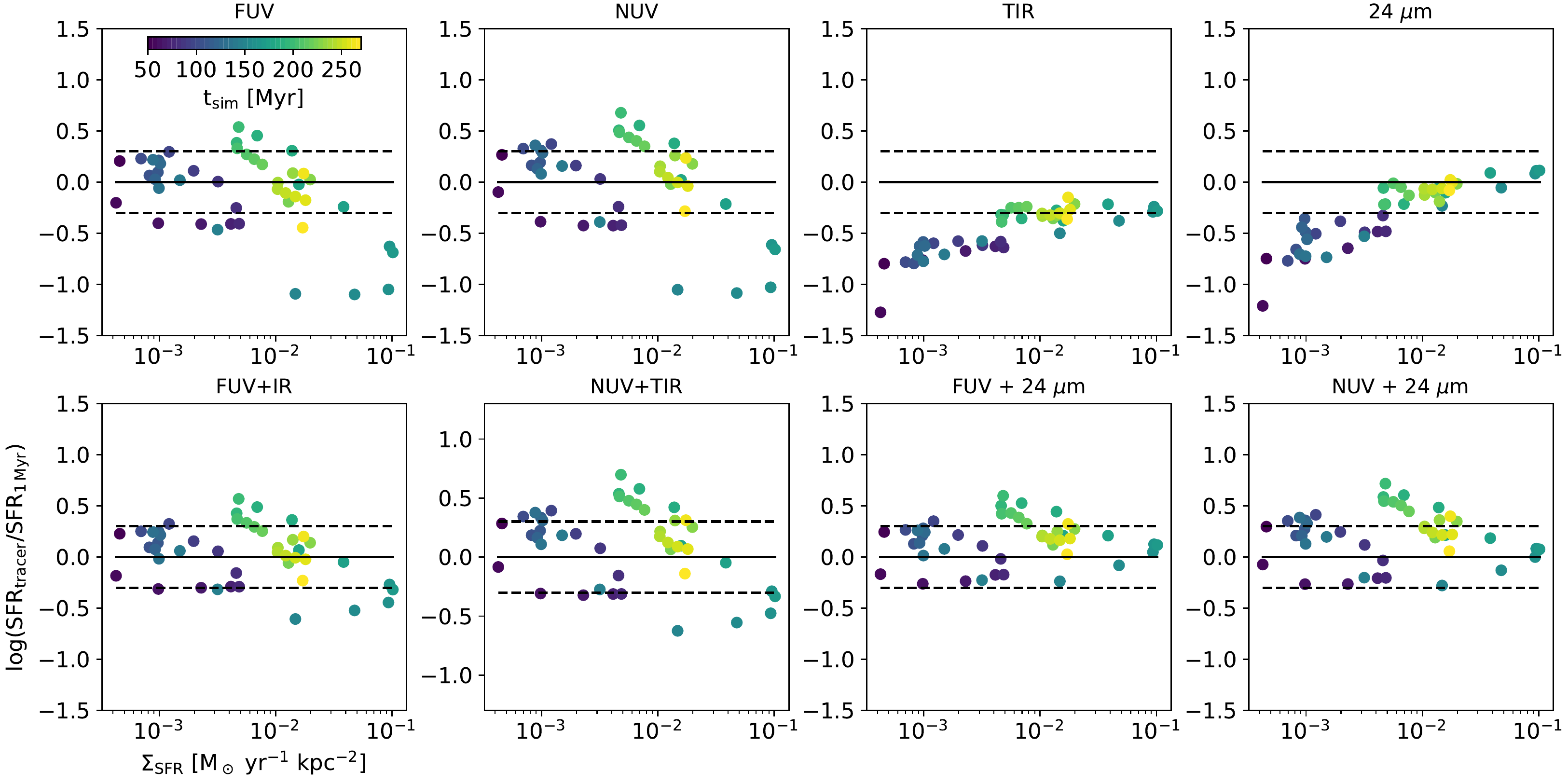}
    \caption{The time evolution of the ratio between the various photometric SFR tracers and the directly integrated SFR, as a function of the SFR surface density ($\Sigma_\mathrm{SFR}$). The particle based SFR is the SFR($\tau=1$ Myr) line from Fig. \ref{fig:SFRs} smoothed to minimize the Myr-to-Myr variations, and the values of $\Sigma_\mathrm{SFR}$ have been calculated in a 100 pc per pixel map for stars younger than 10 Myr. The solid line shows a one-to-one correspondence between the tracer and the underlying smoothed SFR and the dashed lines show the range of a factor of two. The TIR and 24 $\mu$m tracers consistently underestimate the SFR, especially at low $\Sigma_\mathrm{SFR}$, with approximately a power-law relation between the relative offset and $\Sigma_\mathrm{SFR}$, while the UV tracers underestimate the SFR at high $\Sigma_\mathrm{SFR}$ which corresponds to the majority of the starburst. The best agreement with the direct SFR is reached with the 24 $\mu$m corrected UV tracers which for the majority of the merger give values within a factor of two of the underlying result.}
    \label{fig:SFRs_offset}
    \centering
\end{figure*}

The middle panel shows the single-band, dust-affected (except H$\alpha$) SFR estimates while the bottom panel shows the added value of the dust-corrected UV estimates. The UV and IR tracers both underestimate the total SFR at different epochs during the merger depending on how well dust is able to obscure the emitted UV. The total dust mass in this low metallicity system evolves by up to a factor of two of the initial $8\times 10^4$ M$_\odot$. Star formation occurs at lower densities during the more quiescent star formation periods. The IR-based SFRs are for the most parts too low in the quiescent pre-merger stages, as the UV light is able to escape the fairly low density star-forming regions. The most intense star formation periods, such as the onset of the first and second passages as well as the starburst, on the other hand, show more concentrated gas densities to allow the dust to obscure the star formation effectively. Consequently, when star formation ramps up during the first passage and the second encounter/starburst, the UV emission increases but remains embedded until the SFR already starts to decrease after the central blowout. The intrinsic UV flux is up to five times higher than the attenuated UV flux during the starburst. The temporal evolution of the UV  emission translates into a $\sim10$ Myr delay in the UV tracer that is compensated by the rapid response in the IR tracers. The UV emission from massive stars that live for a few tens of Myr also shows as an excess of a factor of 2-3 in the SFR during the first couple tens of Myr after both the first encounter and the starburst, which we discuss below in more detail. The dust-corrected estimates trace the underlying SFR by balancing one tracer with another, and only fail when the slowly fading UV emission remains at an elevated level after the starburst periods. 

We investigate the offset of the SFR estimates in Fig. \ref{fig:SFRs_offset}, where we show the ratio of estimated to direct SFR as a function of star formation surface density along the merger. We adopt as the underlying SFR the 1 Myr curve from Fig. \ref{fig:SFRs} that has been smoothed to minimize the Myr-to-Myr variations but still to follow the SF evolution without the delay introduced e.g. by integrating over the past 10 Myr. The underestimated SFR from the UV emission during the majority of the starburst is quantified in the two top right panels while the underestimated SFRs through the IR tracers are clearly seen to follow a power-law correlation between the offset of the SFR and the SF activity, especially at the low SF activity end of the two top right hand panels. The TIR corrected UV-tracers do not capture the starburst phase either. The shorter wavelength IR corrected UV tracers that recover the direct SFRs to within a factor of two are seen to work best in this setup. We note here that the majority of such SFR tracers have been calibrated at higher, e.g. solar, metallicity, which for IR may especially at low SFRs cause uncertainties in the estimates. With this in mind, the agreement from the FUV + $24\,\mu$m tracer and the underlying true SFR is remarkable.

Finally, the H$\alpha$ tracer in Fig. \ref{fig:SFRs} follows the infrared ones in underestimating the total SFR in more quiescent times as it is as well tied to the distribution of gas. We leave out trying to combine the H$\alpha$ emission with other tracers as we do not include attenuation in its modeling and would therefore over-estimate the correction. It has however been noted in observations of low mass galaxies that the UV bands work better than H$\alpha$ in estimating the SFR when the total star formation activity is low ($\lesssim 0.01$ M$_\odot\mathrm{yr}^{-1}$, \citealt{2009ApJ...706..599L, 2012ApJ...744...44W}), in both original and dust-corrected measurements. The source of the discrepancy is not known, and suggested causes include the calibration not being optimal at lower than solar metallicity, the bursty nature of low mass systems, and incomplete IMF sampling at low SF activity. Our numerical implementation does not allow us to assess this effect in sufficient detail, however see e.g. \citet{2017MNRAS.466.3293P} for a numerical investigation where they found that for example rapid oscillations in the H$\alpha$ flux are caused by the births and deaths of the most massive ($>30$ M$_\odot$) stars, which can lead to underestimation of the SFR by an order of magnitude. Accounting for scattering of H$\alpha$ photons might be able to boost the flux by up to tens of percents  \citep{2022MNRAS.513.2904T} in systems down to Large Magellanic Cloud (LMC) scale, however our low mass system might have too low gas densities to allow for significant scattering. Averaging over a longer timespan (>3 Myr) to mimic a larger system with less stochastic variation would result in H$\alpha$ traced SFR evolution more alike that produced by the infrared tracers, which would essentially still underestimate the underlying SFR for the majority of the simulation time.

Hotter dust emits at shorter wavelengths and is therefore also located closest to the intense star-forming regions. It is thus not surprising that the 24 $\mu$m emission (along with the H$\alpha$) performs better than the composite TIR emission in compensating for rapid variations in SFR evolution, as long as the system is not completely dust-free. When we integrate over the star formation history recovered from the different tracers, the total formed stellar masses compared to the integrated \mbox{SFR$(\tau$ = 1 Myr$)$} are 94\%, 66\%, 85\%, 51\%, 103\%, 93\%, 109\%, 154\% and 162\% in $\mathrm{H}\alpha$, FUV, NUV, TIR, 24 $\mu$m, FUV+TIR, NUV+TIR, FUV+$24\,\mu$m and NUV+$24\,\mu$m. The close to $100\%$ values for $\mathrm{H}\alpha$ and 24 $\mu$m are caused by their ability to catch the starburst, while the FUV+TIR and NUV+TIR values are a combination of underestimated peak SFR and overestimated post-burst SFR that integrated recovers the correct stellar mass. The total stellar mass recovered by the 24 $\mu$m corrected UV calibration on the other hand overestimates the direct value, but best follows the star formation profile in both quiescent and bursty times as seen in Fig. \ref{fig:SFRs_offset}. 

\begin{figure*}
	\includegraphics[width=\textwidth]{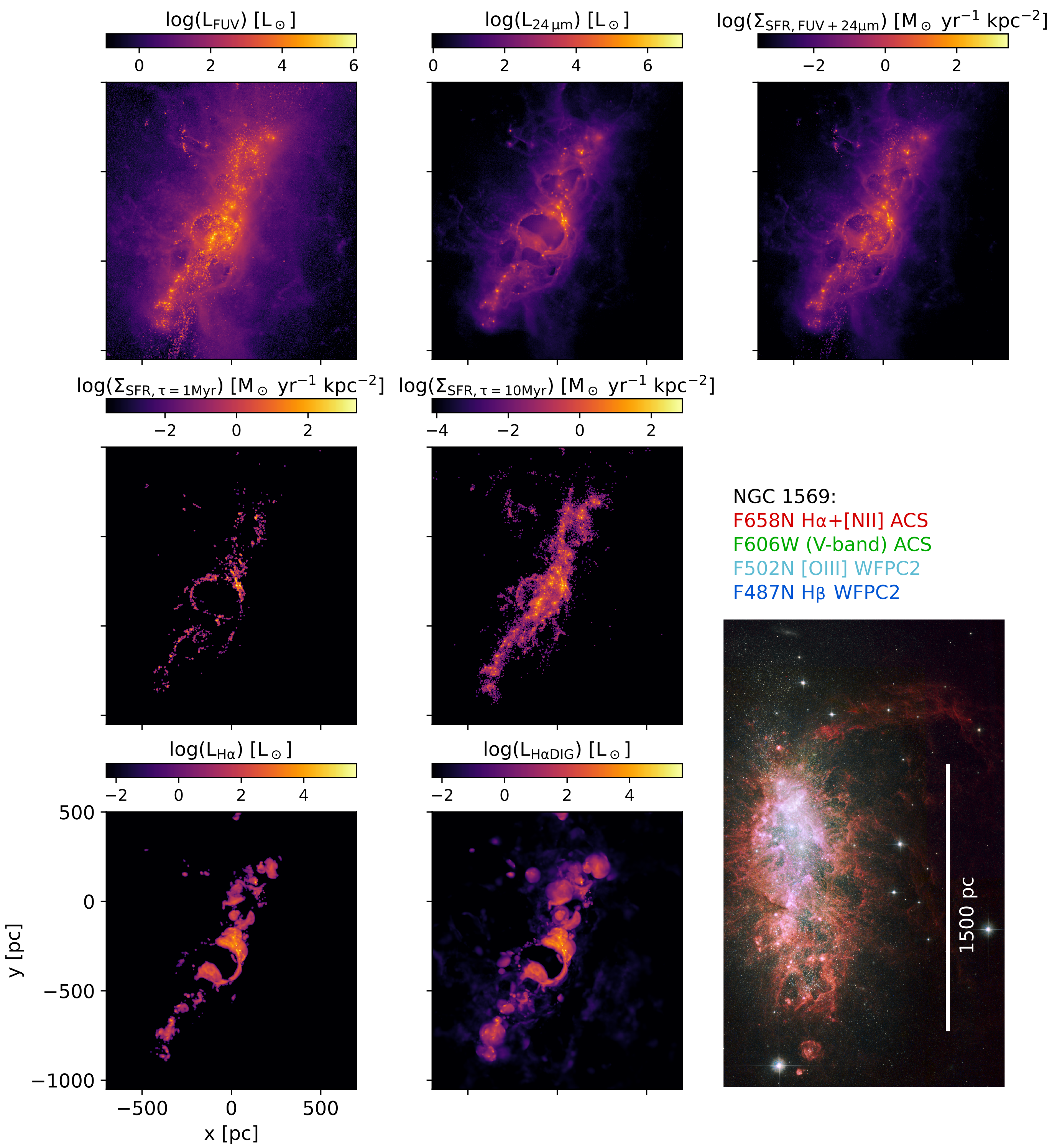}
    \caption{The spatial distribution of FUV, 24 $\mu$m and the star formation rate surface density map estimated from the dust corrected FUV at the time of the most intense starburst (t$_\mathrm{sim}=169$ Myr). The diffuse UV light located further from regions of young stars is mostly scattered light. The SFR tracer maps are compared to the corresponding direct $\Sigma_\mathrm{SFR}$ maps extracted from very young stars \mbox{($<1$ Myr)} and less than 10 Myr old stars. The bottom row shows the H$\alpha$ luminosity map, both excluding (left) and including (middle) the emission from diffuse interstellar gas ($\rho <10$ cm$^{-3}$ or $T>10^{5.5}$ K). The image resolution in the maps is the same as in the top row of Fig. \ref{fig:BVI} and no noise has been added for clarity. The bottom right image shows a \textit{HST} color composite image of NGC 1569, a star-forming dwarf galaxy with similar SFR and gas mass as our merger system, scaled to the same spatial scale. Image credit: NASA, ESA, The Hubble Heritage Team (STScI/AURA), and A. Aloisi (STScI/ESA).}
    \label{fig:spatial_SFRs}
    \centering
\end{figure*}

\subsection{Star formation rate surface density}

In Fig. \ref{fig:spatial_SFRs} we show the spatial distribution of FUV, 24 $\mu$m and H$\alpha$ emission that best trace the evolution of the underlying SFR, as well as the star formation rate surface density ($\Sigma_\mathrm{SFR}$) recovered from the dust corrected FUV emission. As the mapping from luminosity to SFR is linear (as shown in Table \ref{tab:sfr_calibration}), the distribution of emission in each band then directly shows where that tracer would identify star formation. For H$\alpha$, we show in Fig. \ref{fig:spatial_SFRs} the emission that originates predominantly from the dense, $T\sim 10^4$ K gas in HII regions. We also show an H$\alpha$ map that includes the emission from hot diffuse gas ($\rho <10$ cm$^{-3}$ or $T>10^{5.5}$ K), which appears mostly as extended low luminosity bubbles around the star formation regions. For comparison, we show the direct $\Sigma_\mathrm{SFR}$ map extracted both from very young stars ($<1$ Myr) and stars younger than \mbox{10 Myr} that indicate the extent of recent star formation. All the maps have the fiducial 1.5 pc resolution that has been smoothed to the \textit{HST} resolution with a PSF of FWHM=2.1 pix. Note however that this is at least two orders of magnitude better spatial resolution than the Spitzer MIPS and the GALEX instruments would provide at these distances, thus consider for example the IR images here as more alike to what we could expect from the \textit{JWST} MIRI\footnote{https://jwst-docs.stsci.edu/mid-infrared-instrument/miri-instrumentation/miri-filters-and-dispersers} instrument. For simplicity we therefore use the UV and IR filters as proxies to similar future instruments with improved resolution, and assume that all the light is captured in each pixel regardless of resolution and filter. 

As per their nature, the gas/dust tracers (H$\alpha$ and 24 $\mu$m) can only capture obscured or nearby star formation while the UV emission is smeared to larger areas due to scattered emission and migration of the already evolved stars that dominate the spectrum. Regions dominated by young stars and SCs that have blown out their natal gas clouds, such as the cavity blown out by the first massive clusters (see Fig. \ref{fig:BVI}), cannot be enhanced in the IR maps as there is no gas to emit the radiation in such regions. As a result, the $L_\mathrm{FUV}$ map more resembles the $\Sigma_\mathrm{SFR, \tau=10 \,Myr}$ map, while the spatial maps of $L_\mathrm{24\,\mu m}$ and $L_\mathrm{H\alpha}$ look very similar to the $\Sigma_\mathrm{SFR, \tau=1 \,Myr}$ map. The dust corrected FUV map however nicely traces a superposition of the $\Sigma_\mathrm{SFR, \tau=1 \,Myr}$ and $\Sigma_\mathrm{SFR, \tau=10 \,Myr}$ maps. The map shows a variety of main locations of recent star formation, both embedded such as the north-eastern side of the super bubble wall (reminiscent of triggered star formation in observed dwarf galaxies such as Holmberg II, \citealt{2017MNRAS.464.1833E}) and exposed such as SCs in the super bubble. 

We also show a \textit{HST} color composite image of NGC 1569 discussed in Section \ref{section:seds}, that includes H$\alpha$ emission in the narrow-band F658N filter (not corrected for [NII] emission), and the F606W band that is approximately equivalent to the V-band (see Fig. \ref{fig:BVI}). The NGC 1569 image has been matched to the same spatial scale as our images. Comparison of the extent and structure of the H$\alpha$ emission and the visual emission in NGC 1569 to our emission maps and Fig. \ref{fig:BVI} reveals how astoundingly similar the observed and the simulated systems appear. 

\begin{figure*}
	\includegraphics[width=\textwidth]{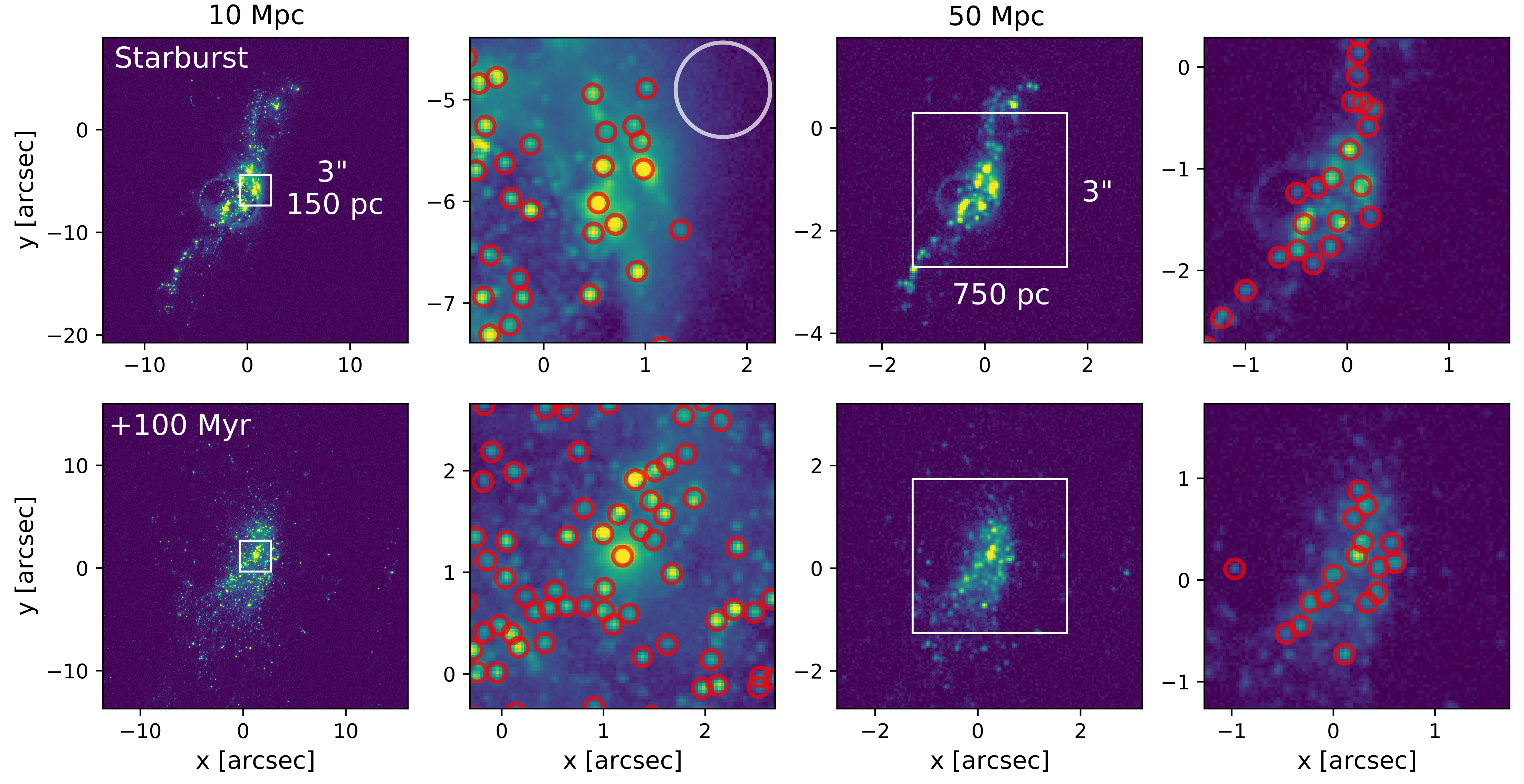}
    \caption{V-band images at the time of the starburst (top row) and 100 Myr later (bottom row). The two left and two right columns show the merger placed at 10 Mpc and 50 Mpc distance, respectively. The $3\arcsec$ zoom-ins in the second and final columns show the final apertures used in the photometry with radii of 3 pixels ($0.093\arcsec\sim 4.5$ pc at 10 Mpc and $\sim 22.5$ pc at 50 Mpc). The images include a \textit{HST}-like noise and PSF, with the color scale adjusted to the noise level of 32 nJy at low brightness end and saturated at higher brightness to bring out also dimmer structures. The images have been centred on the most intense star-forming region (top) and the most massive aperture (bottom). The white aperture in the second panel indicates the 22.5 pc aperture size used in the 50 Mpc images and highlights the crowding in the central star-forming region (top) and the central region of the merger remnant (bottom) that leads to blending in the poor resolution images. }
    \label{fig:apertures}
    \centering
\end{figure*}

\section{Observed star cluster population}

\subsection{V-band selected cluster candidates}\label{section:candidates}

In this section we concentrate on comparing the cluster detection performed at the two resolutions, equal to 10 Mpc and 50 Mpc distances, and at two distinct epochs, i.e. during and significantly after the starburst. Examples of the cluster candidates detected near the most massive clusters at the time of the starburst and 100 Myr later are shown at the two fiducial distances of 10 and 50 Mpc in Fig. \ref{fig:apertures}. The crowding of SCs is obvious around the central star-forming regions during the starburst, but also after the starburst a long time after the clusters have already lost their birth neighborhood. In the good resolution image, dozens of smaller mass clusters populate the central region, whereas in the poor resolution images only the relatively bright clusters can be detected as extended and bright enough in multiple bands as required by the pipeline. It is also evident in these example images how each aperture of the poor resolution image actually contains many small clusters in addition to the brightest, dominating one (see the white circle for a comparison of the aperture sizes at the two resolutions). For a corresponding observed \textit{HST} image, see Fig. 8 of \citet{2013MNRAS.431..554R} where a star-forming region with numerous bright clusters in the Antennae has been translated to a four times lower resolution, blending the clusters together. The effect of crowding is often a challenge for observational cluster catalogues, where apertures that clearly contain many bright peaks are sometimes classified in a different category compared to single clear-cut (circular) SCs when performing for example cluster mass function fits (see e.g. classification in \citealt{2017ApJ...841..131A}). 

\begin{figure*}
	\includegraphics[width=\textwidth]{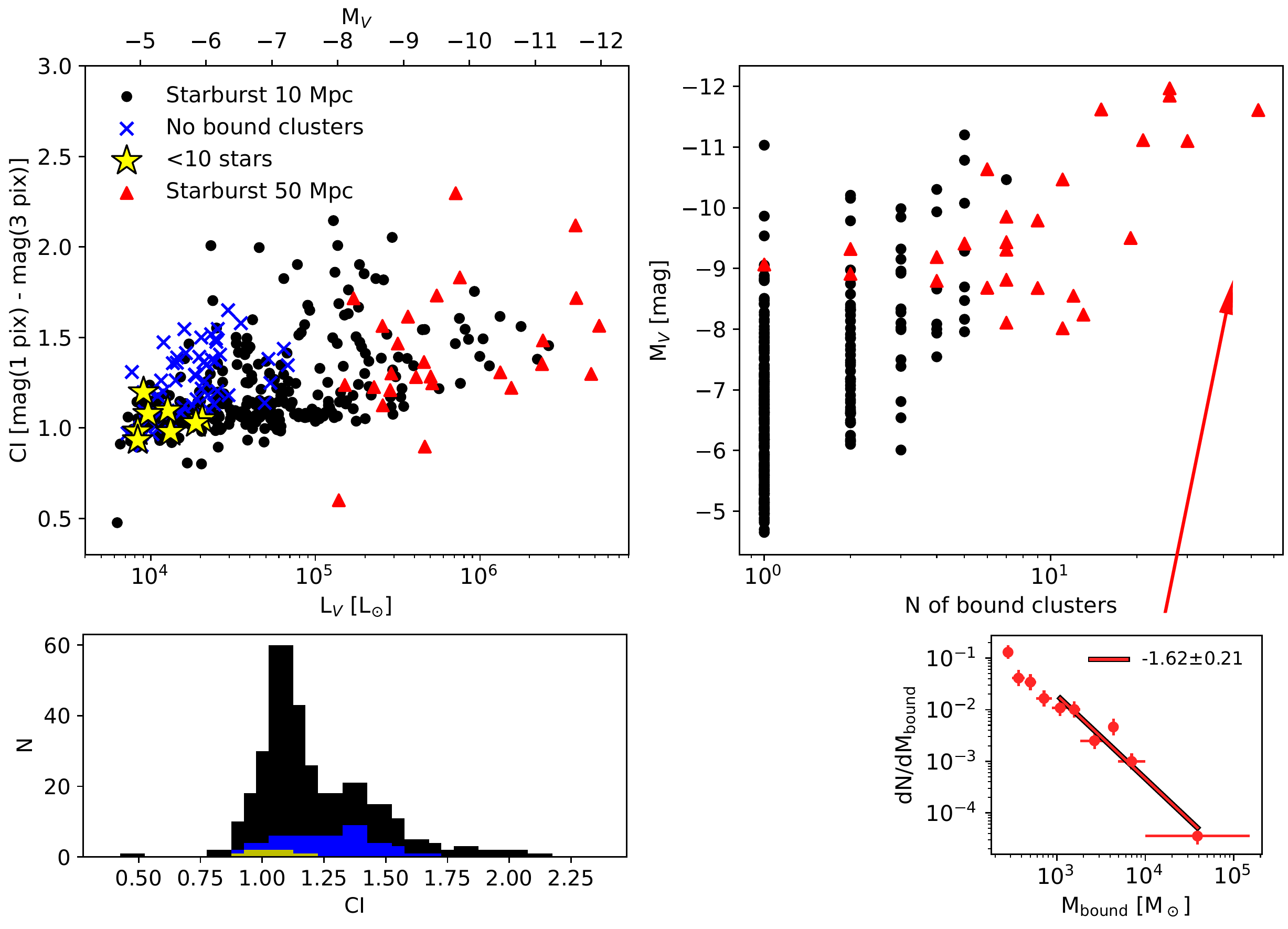}
    \caption{Top left: the concentration index as a function of V-band aperture luminosity and magnitude. The black and red datapoints show all detected apertures when the merger is placed at a distance of 10 Mpc and 50 Mpc, respectively. The blue crosses indicate apertures in the 10 Mpc image that do not enclose any bound clusters ($>200$ M$_\odot$) based on the \textsc{subfind} analysis. Note how many of the apertures that enclose actual bound clusters exhibit quite low values of CI due to the compact size of the stellar distribution of the underlying clusters as discussed in \citet{2020ApJ...891....2L}. Apertures with less than 10 stars (yellow stars) indicate values for bright stars detected by the pipeline. These apertures have CI values between $\sim 0.9$--$1.2$ while in \textit{HST} surveys, stars are typically detected with CI of less than $1.3$--$1.4$. Bottom left: the histogram of the CI values for all detected apertures (black), apertures that do not enclose any bound clusters (blue) and apertures with only a few stars (yellow) shown in the top left. Top right: The integrated V-band magnitude within each aperture that includes at least one truly bound cluster based on the \textsc{subfind} analysis, as a function of the number of bound clusters whose centres of mass fall within each aperture. To highlight the hierarchical nature of the cluster formation process, we show the CMF and the best-fit power-law slope of the 53 bound clusters within the rightmost aperture in the bottom right panel. This corresponds to the most intense star formation region in the system seen in Figures \ref{fig:BVI}, \ref{fig:spatial_SFRs} and \ref{fig:apertures}. Solar V-band magnitude of 4.84 was assumed. }
    \label{fig:CI}
    \centering
\end{figure*}

We investigate the shape and crowding of the clusters in Fig. \ref{fig:CI}, where we show the concentration index (CI) and the number of \textsc{subfind}-identified (i.e. bound) clusters in each aperture. CI is calculated as the difference between the magnitude within a radius of one and a radius of three pixels from the centre of the aperture. Based on our \textsc{subfind} analysis of the cluster population in these snapshots we have also separated the apertures into those that contain one or more truly bound clusters and none.  Apertures which do not include truly bound clusters have in Fig. \ref{fig:CI} wide range of CI values but low luminosity. These are essentially loosely or unbound concentrations of stars, i.e. open clusters and associations, or even single bright stars, as detections of pure noise have been excluded by the multi-band detection criterion. To indicate pure stellar-like CI values, we separate apertures that contain less than 10 stars.

After masking out known foreground and background objects, CI is typically used to exclude foreground stars and bright stars in the observed field (small CI) and background galaxies (large CI) that might be confused for clusters when constructing the initial catalogue of clusters in observed galaxies. Background galaxies, especially ellipticals that may morphologically resemble SCs, can additionally be excluded with cuts in colour. Our simulated observations do not suffer from foreground or background contaminants, but very bright single stars in the field are present in our images. Stars in the \textit{HST} cluster catalogues have often CI$<1.3-1.4$ with some overlap in the values for SCs and even GCs \citep{2010AJ....140...75W}. The histogram in the bottom panel of Fig. \ref{fig:CI} shows how the overall distribution of CI values peaks at $1$--$1.25$ while stellar-like CI values concentrate at CI $\sim1$. The apertures without bound clusters have slightly broader distribution towards higher CI values. Apertures that contain less than 10 stars and have total luminosities of up to $2\times10^4$ L$_\odot$, however, constitute only a few $\%$ of the detected cluster candidates in the advanced stages of the merger, with the fraction increasing towards earlier phases. During and before the time of the first passage, apertures with a few single stars reach a couple tens of $\%$ of the apertures, as only a very few bound clusters are forming (see \citealt{2020ApJ...891....2L}) and single stars can easily outshine the field. 

The apertures that include truly bound clusters in Fig. \ref{fig:CI} however encompass a broad range of CI values and luminosities as well, also within the star-like CI distribution. Our previous analysis of the SC population in the merger \citep{2020ApJ...891....2L} and its progenitor dwarf galaxies \citep{2022MNRAS.509.5938H} have indicated that the SCs have effective radii smaller than the presently used PSFs, which reflects in the more stellar-like values of CI in Fig. \ref{fig:CI}. A cut at $\mathrm{CI}=1.1$ would discard almost $40\%$ of the apertures, including a lot of those that do indeed include real bound clusters. We therefore resort to excluding the stellar-like apertures based on the number of stars rather than using a cut in CI. The cluster candidates in the low resolution (50 Mpc) image populate higher luminosities as the majority of the low luminosity objects are drowned by noise. None of the apertures in the poor resolution image are single star contaminants. At a distance of 50 Mpc, even massive SCs appear point-like, and the CI indicator becomes less reliable \citep{2020MNRAS.499.3267A}.

The right panel of Fig. \ref{fig:CI} shows how the brighter regions also suffer from worse crowding. As discussed in our earlier studies, the formation of the most massive clusters is hierarchical. In regions that host young SCs, each massive (proto)cluster is surrounded by numerous smaller mass clusters. In bright apertures, the light is therefore actually composite from multiple individual bound clusters even though the brightest cluster may in principle dominate the luminosity. The effect here is stronger in the poor resolution images, where one of the brightest apertures includes more than $50$ individually bound objects. The inset in the right panel of Fig. \ref{fig:CI} shows the binned CMF of 53 individual bound clusters in the most intense star formation region in the merger, which corresponds to the obscured region indicated in the left hand panel of Fig. \ref{fig:BVI}. 
Results of similar nature have been discussed in \citet{2010AJ....140...75W} where the authors looked at the number of clusters in regions of varying star formation activity. Regions with the most intense star formation in the Antennae merger harbour the brightest individual clusters but also host the largest number of detected SCs. 
We can fit power-law mass functions with slopes between $\sim -1.6$ and $-1.9$ to the bound clusters ($>10^3$ M$_\odot$) contained within the four of our brightest poor-resolution apertures, of which an example is in the bottom right panel of Fig. \ref{fig:CI}. The slopes are of similar order as the most intense star formation regions in the Antennae that have slopes as shallow as $-1.6$, and demonstrate how the cluster formation process in the high mass end of the CMF is hierarchical. The masses of the SSCs interpreted through photometry of starbursting environments, especially at large distances, should therefore be viewed as upper limits when inferring to them as possible proto-GCs. Distant young SSCs observed in surveys such as \citet{2021ApJ...912...89K} with traditional aperture photometry using aperture radii of tens or hundreds of parsecs ($\gg$ half-light radii) contain very unlikely only single clusters. \textit{JWST} and strong lensing studies with e.g. \textit{HST} and VLT/MUSE provide great tools to access better resolved studies of more distant star-forming structures

\begin{figure}
	\includegraphics[width=\columnwidth]{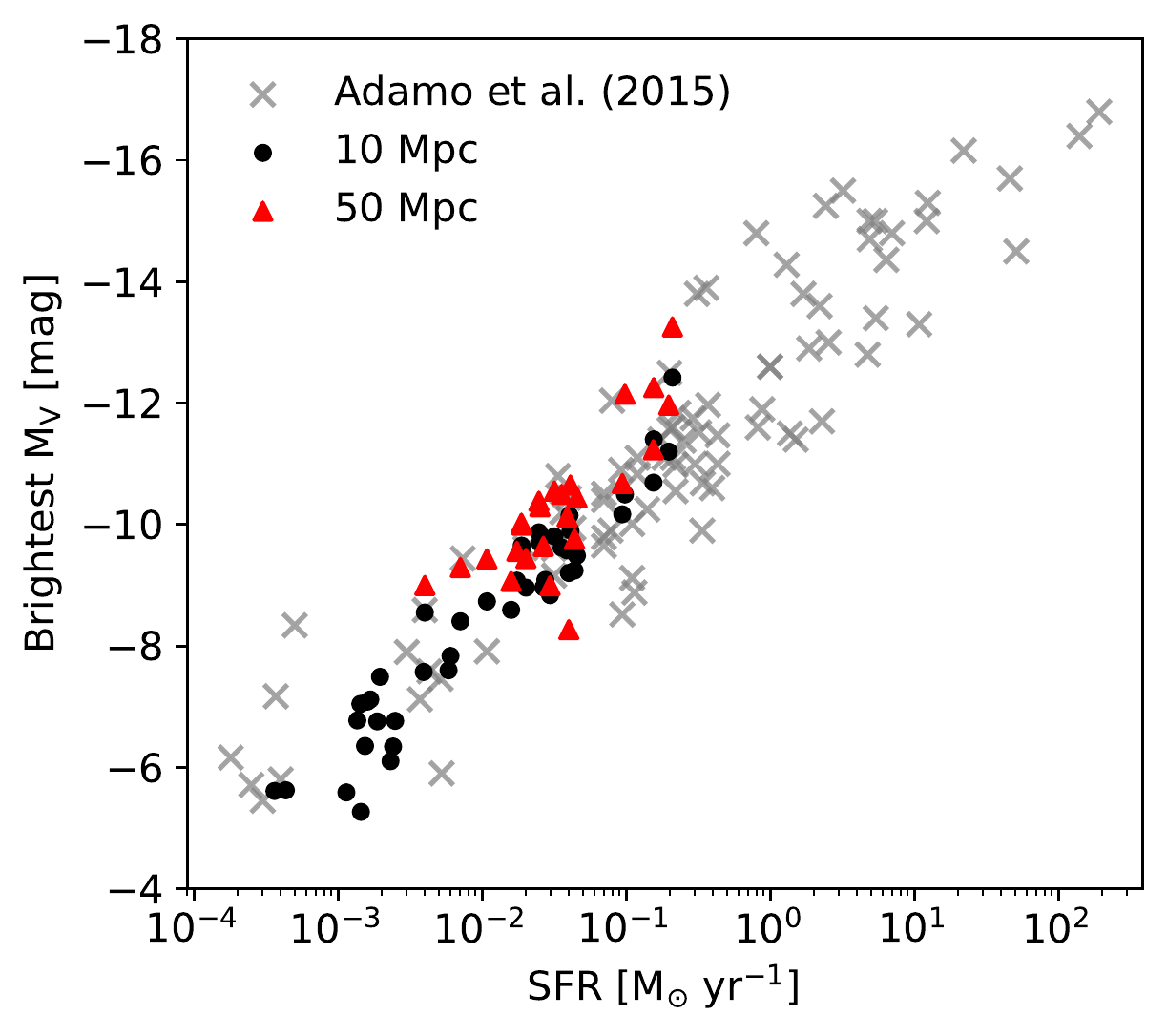}
    \caption{The brightest V-band magnitude of the photometrically detected clusters as a function of star formation rate. Each point shows the brightest young cluster ($<10$ Myr) of a single snapshot (with 5 Myr interval) using the resolution at 10 Mpc (black circles) and 50 Mpc (red triangles). Only apertures that include at least 10 stars are considered. The snapshots with lowest SFRs that correspond to early stages in the merger do not have any detections at 50 Mpc distance. The SFR is estimated from all stars younger than 10 Myr in each snapshot, and the value is the same for both resolutions. Observed data shown with gray crosses, that includes a variety of galaxies such as nearby dwarfs, luminous blue compact galaxies and major mergers, are from Table B1 of \citet{2015MNRAS.452..246A}.}
    \label{fig:MV_SFR}
    \centering
\end{figure}

In \citet{2019ApJ...879L..18L} we verified that the mass of the most massive SC depends on the star formation environment as expect from observations (see e.g. \citealt{2002AJ....124.1393L,  2008MNRAS.390..759B, 2015MNRAS.452..246A, 2017ApJ...839...78J}), theory \citep{2017MNRAS.469.1282R, 2018ApJ...869..119E} and cosmological simulations (\citealt{2017ApJ...834...69L, 2019MNRAS.490.1714P} where the latter implements the analytical model of \citealt{2017MNRAS.469.1282R}). Unlike cosmological simulations that often resort to sub-resolution models to follow the SC populations, here we self-consistently recover the maximum cluster mass by following the resolved collapse and accretion of gas and stars into bound structures. Highest mass clusters form in our simulation in environments that experience largest values of SFR and $\Sigma_\mathrm{SFR}$. In Fig. \ref{fig:MV_SFR} we show the V-band magnitude of the brightest aperture with luminosity-weighted age less than 10 Myr in each snapshot along the entire merger sequence (in 5 Myr steps). The low SFR snapshots are missing from the 50 Mpc data as at that distance, the detection pipeline does not recover any clusters in the early stages of the merger. The reference data collected in \citet{2015MNRAS.452..246A} is shown on the background. The aperture magnitudes agree well with observed values that also include dwarf galaxies and luminous blue compact galaxies. Interestingly, integrating within a larger aperture of our 50 Mpc resolution images gives still values in reasonable agreement with the observed data. Based on Fig. \ref{fig:CI} we know the brightest apertures in the 50 Mpc images can actually contain tens of bound clusters.

\subsection{Cluster mass and luminosity function}\label{section:CMF}

Instead of adding up uncertainties through attempting to fit the cluster mass in each aperture with, for example, SSP models, we extract the stellar mass in the apertures directly from the stellar densities given by the particle data in the simulation snapshots. We construct the surface density maps by projecting the stellar particle masses onto a map with equivalent resolution as the corresponding photometric image, and integrate within the same photometrically selected aperture radii. The aperture integrated cluster masses are corrected for the background, the sky annulus estimate and  the median aperture correction in a similar way as in the photometry but using the stellar density map instead. The SCs in our simulated system are still somewhat lower mass objects compared to the SSCs in excess of $10^6-10^7$ M$_\odot$ in some of the most extreme starbursting galaxies \citep{2010AJ....140...75W, 2020MNRAS.499.3267A, 2021ApJ...912...89K}. As discussed in the previous paragraphs, our clusters tend to also be fairly compact in size compared to the image resolution. The largest half-mass radii during the starburst reach 4.5 pc and even 100 Myr later do not exceed 10 pc. The median aperture correction for mass in the 10 Mpc images are therefore typically $20 \%$. As with the photometry, we only correct the 10 Mpc images due to lack of reasonably isolated bright enough clusters in the 50 Mpc images.

We then use the aperture integrated luminosities and stellar masses to fit luminosity and mass functions of the form $dN/dL\propto L^{\alpha_L}$ and $dN/dM\propto M^{\alpha_M}$ with power-law indices $\alpha_L$ and $\alpha_M$. These fits represent the photometrically selected cluster mass and cluster luminosity functions (CMF, CLF). Additionally, we investigate the distribution of truly bound clusters that fall within each aperture, in order to compare the loss of information in the CMF through crowding (many clusters in single aperture) versus brightness (low brightness clusters indistinguishable from noise). 

\begin{figure*}
	\includegraphics[width=\textwidth]{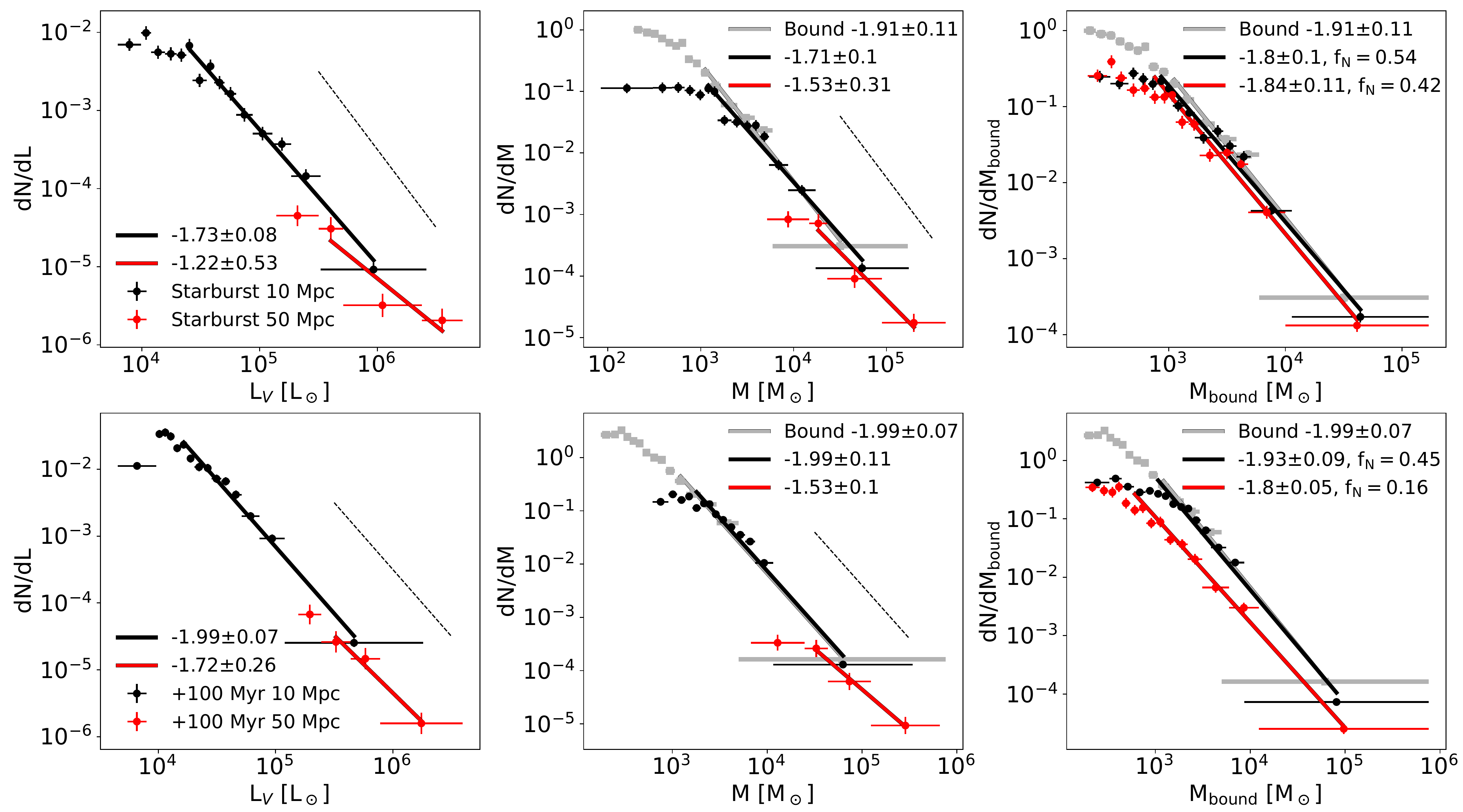}
    \caption{The final cluster V-band luminosity (left) and mass (middle) functions obtained through aperture photometry. Apertures with less than ten stars have been excluded. We show the binned data at the two epochs (starburst and 100 Myr later, top and bottom rows respectively) and two image resolutions (equivalent to 10 Mpc and 50 Mpc distance, black and red data points). The best fit power-law slopes (equivalent to $\alpha_L$ and $\alpha_M$) are shown with same colors as the datapoints (see text for details). The rightmost panel shows the underlying CMFs based on all \textsc{subfind} selected truly bound clusters (each bound cluster with mass M$_\mathrm{bound}$) that have their centre of mass within any photometrically selected aperture. The fraction of captured bound clusters by number is indicated in the legend. The full, bound CMF at both epochs is shown as the gray squares and the best fit line on the background of the middle and right hand panels. The data has been divided in order of decreasing luminosity/mass into bins that include equal number of apertures/clusters, with 15 (bound clusters and 10 Mpc data) or 4 (50 Mpc data) bins. The horizontal errorbars show the bin widths, which also indicate the maximum and minimum values of the aperture data and the bound cluster masses along each x-axis, and the vertical errorbars show the Poisson error. The thin dashed line indicates a -2 power-law slope.}
    \label{fig:CMFs}
    \centering
\end{figure*}

We show the various fits and the underlying binned data in Fig. \ref{fig:CMFs}. The left and middle panels show the luminosity and mass functions based on the final set of V-band selected apertures at the time of the starburst (top) and 100 Myr later (bottom). We only show apertures that include at least ten stars (see Fig. \ref{fig:CI}). The right hand panel shows the CMF of the individual, truly bound \textsc{subfind} selected clusters captured within the apertures as discussed in relation to Fig \ref{fig:CI}. The power-law fits have been done to bins as follows. The bound cluster data (gray data points) and the data from the 10 Mpc images (black data points) has been binned into 15 bins of equal number of clusters/apertures per bin. The CMF fit of the bound clusters is done to $>10^3$ M$_\odot$ bins. The slope of the 10 Mpc data is always fit to the 10 highest mass/luminosity bins. By inspecting the binned data from the starburst onward, we verified that the simple power-law shape is mostly retained by these bins, as one can examine in the black lines of Fig. \ref{fig:CMFs}. In the 50 Mpc images we only recover the 20--30 brightest clusters. Here we divide the data into four bins and fit the power-law to the three highest mass/luminosity bins. Due to the low number of clusters, the power-law fits are somewhat uncertain and one should therefore mostly concentrate on the actual data points with respect to the 50 Mpc results. The power-law index of each single power-law fit and the number fraction of bound clusters captured in the apertures are indicated in the legend of the right hand panels. The fits shown in the two left columns of Fig. \ref{fig:CMFs} include also the apertures where we know that no actual bound clusters reside. If we leave these typically low luminosity cluster candidates out, the best fit power-law indices change very little, only by $\sim0.02$ or less. 

\begin{figure}
	\includegraphics[width=\columnwidth]{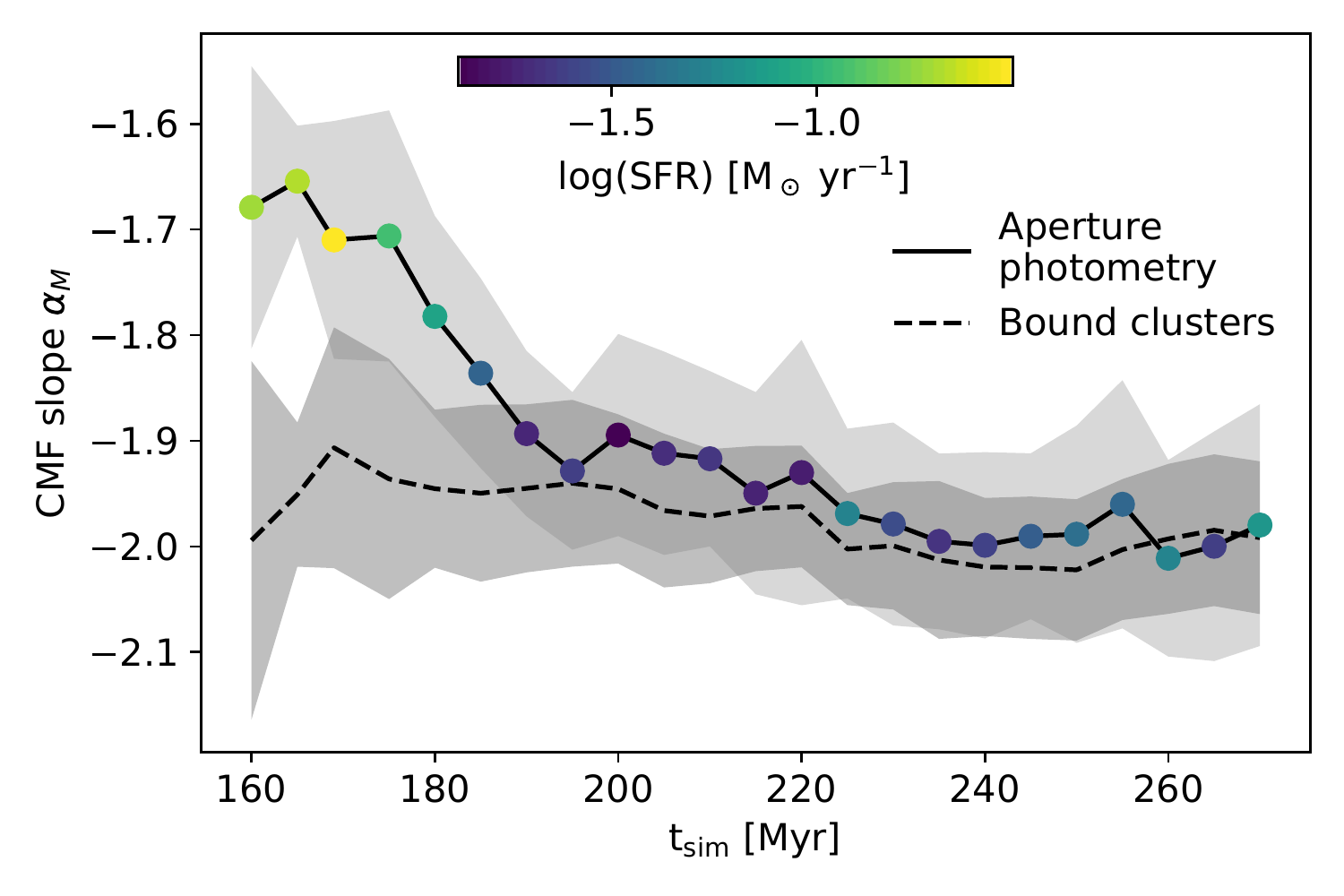}
    \caption{The evolution of the best-fit power-law indices of the photometrically selected final cluster population (solid line with markers) in the 10 Mpc (1.5 pc per pixel) images and in the underlying bound SC population (dashed line). The colour-coding of the markers indicate SFR to show how the shallowest CMF slopes coincide with the starburst. The fits have been done to binned data as in Fig. \ref{fig:CMFs} and the shaded regions show the standard error of the power-law index. The recovered slopes are shown from \mbox{$t=160$ Myr} onward, when the simulation samples the CMF beyond $10^5$ M$_\odot$.}
    \label{fig:CMF_slopes}
    \centering
\end{figure}

The first immediate result of Fig. \ref{fig:CMFs} is that the power-law indices of the fits based on the photometric analysis obtained at the time of the starburst (top row) deviate from the underlying bound CMF at both image resolutions. This is for the most parts due to crowding, as can be seen by comparing to the corresponding data in the top right panel. The bound cluster distribution that is captured within the apertures (black and red data on the right) shows nearly the same power-law as the corresponding cluster data (gray), with a 0.1 shallower slope and a slightly lower overall normalization due to incompleteness especially at lower masses. When performing aperture photometry, light from the increasingly numerous low-mass clusters that surround the bright clusters is simply blended together, as was indicated by the right hand panel of Fig. \ref{fig:CI}. This leads to transfer of mass (light) from the low-mass (low-luminosity) end of the CMF (CLF) to the high-mass (high-luminosity) end in the left and middle panels of Fig. \ref{fig:CMFs}. The very uncertain slopes of the photometric 50 Mpc data (left and middle panels) are considerably shallower than the underlying data. A similar study with a bigger sample and observed clusters were studied by \citet{2013MNRAS.431..554R} who investigated the effect of worsened resolution by transforming \textit{HST} observations of the Antennae galaxies to a four times greater distance. They found that typically the subsequent blending should not affect the LF power-law by more than $0.05$--$0.1$. Low number statistics does not allow us to attempt similar conclusions, however the power-law mass/light functions are not completely erased even when the recovered sample only covers the most massive/bright clusters.

We take a closer look at the snapshot-to-snapshot variation of the power-law indices in the 10 Mpc images across and past the starburst in Fig. \ref{fig:CMF_slopes}. Here we compare the photometrically recovered best-fit CMF power-law indices to the CMF slope of the bound cluster population in 5 Myr steps. The slopes are shown for times when the CMF is filled upwards of $10^5$ M$_\odot$, i.e. from $t=160$ onward. During the starburst, the recovered slope in the good resolution images is $\sim0.2$--$0.3$ too shallow. After the starburst, once clusters formed during the starburst have expelled and left their gas-rich environments and had time to dynamically separate from their birth neighbours, the photometric detection pipeline performs somewhat better. After the starburst the slope of the bound SC population evolves only slightly from $-1.9$ to $-2.0$ ($\pm0.1$) while the slope recovered from photometry reaches $-1.9$ approximately 20 Myr after the peak starburst. Afterwards the photometric slope follows quite closely the true slope, especially taking into account the standard errors of the best-fit values.

\begin{figure*}
	\includegraphics[width=\textwidth]{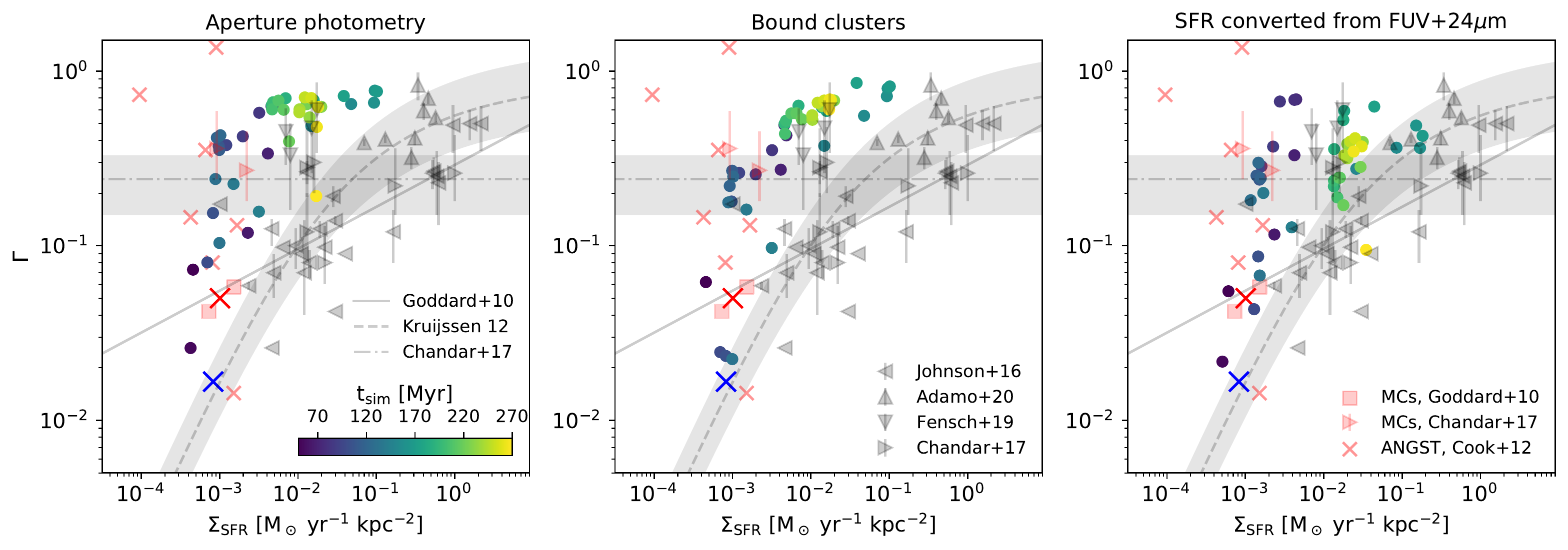}
    \caption{The cluster formation efficiency across the dwarf galaxy merger in 5 Myr steps. The left panel shows the CFE based on stellar mass in apertures with luminosity weighted mean age younger than \mbox{$<10$ Myr} and at least 10 stars, while the middle panel shows the CFE in the young bound SC population from \citet{2020ApJ...891....2L}. The right hand panel shows the same as on the left but using the SFR recovered using the 24 $\mu$m corrected FUV tracer as shown in Fig. \ref{fig:SFRs}. The observational reference data are from  \citet{2010MNRAS.405..857G}, \citet{2012ApJ...751..100C}, \citeauthor{2016ApJ...827...33J} (\citeyear{2016ApJ...827...33J}, Table 5 and references therein), \citet{2017ApJ...849..128C}, \citet{2019A&A...628A..60F} and \citet{2020MNRAS.499.3267A} (HiPEEC) with dwarf galaxies in the ANGST survey and the Magellanic clouds highlighted in red. The big crosses show the averaged values for all dwarfs in \citet{2012ApJ...751..100C}, considering only young clusters ($<10$ Myr, red) and also including older clusters ($<100$ Myr, blue). The best fit relation from \citet{2010MNRAS.405..857G} is shown with a solid line, the analytic model from \citet{2012MNRAS.426.3008K} with a dashed line along with a 0.2 dex range in gray background, and the best fit constant value of $24\%\pm9$ from \citet{2017ApJ...849..128C} is shown with dot-dashed line.}
    \label{fig:gamma}
    \centering
\end{figure*}

Overall, all the power-law indices recovered both from the aperture luminosities ($\alpha_L$) and from the direct integration of the stellar surface density maps ($\alpha_M$) are the same or shallower compared to the underlying bound cluster population, but differ most drastically at the peak starburst. As discussed in relation to Fig. \ref{fig:CI}, similar effect of blending may be taking place in the Antennae merger analysis of \citet{2010AJ....140...75W} as well, where the cluster luminosity and mass functions in regions of more intense, recent star formation activity (young ages and more central locations) have slightly shallower power-law indices up to $-1.6$, compared to very outermost, older regions with indices only steeper than $-1.9$. The merging disk galaxies in HiPEEC, on the other hand, show little to no difference between young and aged cluster populations, while their early stage mergers (before the coalescence of the nuclei) show steeper slopes than advanced mergers (single disturbed nucleus with strong or weak tidal tails). They report shallower than $-2$ power-law indices across the board (mean values from $-1.60$ to $-1.86$, \citealt{2020MNRAS.499.3267A}), even though they leave out of the analysis all clumpy objects and the central knots. Shallower mass function indices of star-forming regions with increasing SFR have been reported in \citet{2016MNRAS.462.3766C} who, like \citet{2013MNRAS.431..554R} for SCs, do not find significant effect of blending on the power-law slope.

\citet{2019MNRAS.490.1714P} investigated the power-law index - SFR relation in cosmological simulations and found a similar correlation as \citet{2016MNRAS.462.3766C}. Notably, \citet{2019MNRAS.490.1714P} did not simulate blending effects, rather the reduced number of low mass clusters was explained through their increased destruction in high SFR environments. In Fig. \ref{fig:CMF_slopes} we show that the shallowest best-fit slope, both based on bound clusters and photometry, coincides with the starburst. The effect is pronounced in the photometrically recovered slope that is statistically different from the slope of the bound cluster population during the starburst. The slope of the bound clusters, on the other hand, does not evolve significantly. The shallower photometric mass slope during the starburst is therefore caused by photometric effects, namely crowding.

In Fig. \ref{fig:CMFs}, the horizontal bars show the extent of each bin and consequently indicates the highest recovered cluster mass in each image. Based on the \textsc{subfind} analysis, there are four intrinsically bound clusters between masses of \mbox{$1.01\times10^5$ M$_\odot$} and \mbox{$1.69\times 10^5$ M$_\odot$} in the starburst snapshot. The photometric pipeline recovers three clusters with aperture corrected masses between \mbox{$1.10\times10^5$ M$_\odot$} and \mbox{$1.74\times 10^5$ M$_\odot$} in the \mbox{10 Mpc} image (1.5 pc per pixel). The most massive clusters, when matched to their bound counterparts, have correct masses to within $-34\%$ and $+15\%$. In the 50 Mpc (7.5 pc per pixel) image the pipeline recovers five clusters between \mbox{$1.24\times10^5$ M$_\odot$} and \mbox{$4.41\times 10^5$ M$_\odot$}. For the massive clusters, the extracted cluster masses overestimate the respective bound cluster mass by few tens of percent and up to a factor of $2.9$. This is due to blending of nearby stars and clusters in the very crowded cluster formation regions.

\citet{2013MNRAS.431..554R} also investigated the effect of worsened resolution on the aperture integrated cluster properties. They found that the  majority of the less bright detections were affected not more than 0.1 mag when the image resolution was made worse by a factor of four. Clusters in the brightest star-forming knots, however, blended together, in a very similar fashion as our starburst regions. As a result they recovered for example multiple objects that were roughly 0.4 mag and one case that was 1.1 mag brighter than in the objects in the original images, which correspond to an increase in brightness of $45\%$ and $175\%$. These are very similar results to our recovered clusters at the two distances, where the most massive clusters are detected with $20$--$190\%$ higher masses in the five times worse resolution starburst image. 

Finally, looking again at the highest mass clusters, now \mbox{100 Myr} after the starburst, there are two truly bound clusters more massive than \mbox{$10^5$ M$_\odot$} with total bound masses of \mbox{$1.20\times 10^5$ M$_\odot$} and \mbox{$7.61\times 10^5$ M$_\odot$} based on \textsc{subfind} analysis. The good resolution photometry provides one cluster above \mbox{$10^5$ M$_\odot$}, with aperture mass of \mbox{$3.43\times 10^5$ M$_\odot$}. The second most massive cluster has \mbox{$8.75\times 10^4$ M$_\odot$}. These most massive clusters here are therefore recovered with only $45$--$73\%$ of the total bound mass.  The most massive clusters are the only few clusters that have half-mass radii close to the aperture size. The recovered mass could therefore be improved by modelling the mass profile in more detail instead of using the median aperture correction, in order to recover the lost mass in the outskirts. The poor resolution analysis of the post-starburst dwarf recovers five clusters with masses in excess of $10^5$ M$_\odot$. The most massive cluster is detected with a mass of $6.62\times 10^5$ M$_\odot$, only $13\%$ below the true bound mass.

\subsection{Cluster formation efficiency}

In \citet{2020ApJ...891....2L} we discussed the efficiency of star formation that happens in bound clusters, based on the \textsc{subfind} analysis of the snapshot data from the same simulations discussed here. This cluster formation efficiency (CFE or $\Gamma$) is often defined either as the ratio between cluster mass $M_{\mathrm{cl}}(<\tau)$ and stellar mass $M_{*}(<\tau)$ formed across the same time interval ($\tau$), or as the ratio between the cluster formation rate (CFR) and star formation rate as
\begin{equation}\label{eq:gamma}
    \Gamma = \frac{M_{\mathrm{cl}}(<\tau)}{M_*(<\tau)} \, \mathrm{or} \, \frac{\mathrm{CFR}_{\mathrm{cl}}(<\tau)}{\mathrm{SFR}  (<\tau)}.
\end{equation}
The time interval considered for recent cluster formation is often within the most recent 10 or 100 Myr. Our earlier analysis recovered values from a few \% to 90 \% for $\tau=10$ Myr across two order of magnitude in the star formation rate surface density with positive correlation between $\Sigma_\mathrm{SFR}$ and CFE.

Here we approximate the CFE by calculating the stellar mass in the final photometric apertures that have surface brightness weighted mean stellar ages of less than 10 Myr. This is very inclusive and does not consider any boundness or compactness arguments. We exclude apertures that include less than 10 stars, which namely affects datapoints in the early merger stages (before second encounter). The star formation rate surface density is estimated in a galaxy-wide grid with a 100 pc pixel scale and pixels with no star formation are excluded when calculating the global mean. 

In Fig. \ref{fig:gamma} we compare the present results to the CFE values obtained from the bound SC population, as well as to a set of observed values for a range of galaxies from dwarfs to massive starbursts. The left hand panel shows the analysis based on the good resolution (10 Mpc distance, 1.5 pc per pixel) photometric images performed here, and the middle panel repeats the results for all bound clusters younger than 10 Myr from \citet{2020ApJ...891....2L}. The right hand panel shows the CFE based on aperture photometry but using the SFR converted from the 24$\mu$m corrected FUV (see Fig. \ref{fig:SFRs}, table \ref{tab:sfr_calibration}) instead of the true underlying SFR to estimate $\Sigma_\mathrm{SFR}$ and the total young stellar mass in Eq. \ref{eq:gamma}. The surface area given by the true SFR map is used to convert the extracted SFR into $\Sigma_\mathrm{SFR}$. We use here the good resolution images which capture fairly well the properties of the cluster population as discussed in Section \ref{section:CMF}, unlike the poor resolution images where the detection only recovers the high mass end of the CMF.

In the observational reference data, the highest CFE values are found in merging disk galaxies \citep{2020MNRAS.499.3267A} while the highest $\Sigma_\mathrm{SFR}$ values are interestingly found in blue compact dwarf galaxies \citep{2011MNRAS.417.1904A}. The majority of the datapoints at the high $\Sigma_\mathrm{SFR}$ end are results based on clusters younger than a few tens of Myrs. The \citet{2019A&A...628A..60F} data are for clusters younger than 30 Myr, harboured in tidal dwarf galaxies, while the \citet{2017ApJ...849..128C} and the \citet{2020MNRAS.499.3267A} data strictly only consider $<10$ Myr old clusters. The CFE measurements of clusters selected solely based on young age are inherently inclusive, as they include all stellar concentrations whether they are bound or not. This corresponds to the way we have processed our snapshots in the left hand panel of Fig. \ref{fig:gamma}. As a result, the very inclusive studies seem to argue for higher CFE values \citep{2011MNRAS.417.1904A, 2020MNRAS.499.3267A} that may even be independent of $\Sigma_\mathrm{SFR}$, such as the constant CFE of $24\%$ by \citet{2017ApJ...849..128C}. CFE studies that have an emphasis on bound clusters, on the other hand, can be used more reliably even in regions with low SF activity as they often include a broader range of cluster ages. Such observations naturally find lower values of CFE \citep{2008MNRAS.390..759B, 2011A&A...529A..25S, 2016ApJ...827...33J} compared to the inclusive methodology at corresponding $\Sigma_\mathrm{SFR}$ values, but with a stronger evidence for a CFE--$\Sigma_\mathrm{SFR}$ correlation \citep{2010MNRAS.405..857G} across a wide range of $\Sigma_\mathrm{SFR}$. Our bound cluster analysis is more alike this latter methodology, as we essentially capture the same bound clusters that would survive for longer periods of time. We however only include them until the age of 10 Myr in order to follow the variation more closely, allowed by us knowing our bound sample is complete at all times. The middle panel of Fig. \ref{fig:gamma} correspond thus to the more exclusive methodology, as well as the analytic disk model for bound cluster formation from \citet{2012MNRAS.426.3008K}. 

Consequently, the CFE values for dwarf galaxies span a wide range that results from stochasticity of the observed CMF and the true dwarf-to-dwarf variation \citep{2012ApJ...751..100C}, as well as the specific definition of CFE discussed above. Starting with the Magellanic clouds, \citet{2010MNRAS.405..857G} datapoints for LMC and the Small Magellanic Cloud include clusters up to ages of $100$ Myr, which results in CFE values lower by multiple factors compared to the corresponding \citet{2017ApJ...849..128C} results. In the ANGST\footnote{ACS Nearby Galaxy Survey Treasury} dwarf galaxy sample of \citet{2012ApJ...751..100C} the CFEs for individual dwarfs with $\Sigma_\mathrm{SFR}<10^{-2.5}$ span from a few $\%$ up to more than $100\%$. Note here that for the CFE defined as CFR per SFR, a varying star formation history probed by different observational quantities (e.g. aperture photometry vs. H$\alpha$ tracer) may give CFE values not limited to $0$--$100\%$. We show in Fig. \ref{fig:gamma} only the individual CFE measurements based on young clusters ($<10$ Myr) in dwarf galaxies that have certain cluster detections, but show the averaged results for the full sample of both only young ($<10$ Myr) and $<100$ Myr old clusters. As with the Magellanic clouds, the strictly young population in ANGST has a three times higher mean CFE compared to the sample which includes also older clusters. Most interestingly, the trend in the ANGST data for the CFE and $\Sigma_\mathrm{SFR}$ is inverse to the results in higher mass galaxies and larger $\Sigma_\mathrm{SFR}$ values, such as depicted by the best fit $\Gamma\propto\Sigma_\mathrm{SFR}^{0.24}$ relation of \citet{2010MNRAS.405..857G}. The lower star formation activity dwarfs seem to show higher fractions of clustered star formation, compared to the more actively star-forming dwarf galaxies, accompanied with a wide range of scatter. Some of the ANGST galaxies were revisited  recently by \citet{2019MNRAS.484.4897C}, who found that the cluster mass measurements supplemented by ground based fluxes may be overestimated due to crowding. The CFE values for the ANGST galaxies may therefore be upper limits.

A comparison between our data in the left and the middle panels shows how the inclusive photometric selection recovers fairly similar values of CFE compared to the bound young clusters during and after the starburst. Only before the second passage are the recovered photometric CFE values somewhat higher than what the fraction of bound clusters provides. This happens because at times when only a few lower mass bound clusters are forming, the photometrically detected population is dominated by associations and open clusters that are either not bound or have very low masses and therefore do not appear in the \textsc{subfind}-based data. The CFE values at early phases of the merger are typically 20--80$\%$ or so higher in the inclusive analysis compared to the full young cluster population. The largest differences by a factor of 2--6 coincide with the first minimum of bound cluster formation rate after the first passage (see Fig. 5 in \citealt{2020ApJ...891....2L}). 

The final results of our photometric analysis coincide with the \citet{2012ApJ...751..100C} results in the low $\Sigma_\mathrm{SFR}$ end while the active star formation periods during the starburst are closer to CFE $\sim$70--80$\%$, and thus the results of \citet{2019A&A...628A..60F} and \citet{2020MNRAS.499.3267A}. These observations use only young clusters in recovering the value of CFE. On the other hand, the analysis based on the bound clusters follows slightly better the trend in the analytic relation of \citet{2012MNRAS.426.3008K} and the overall trends seen in higher mass galaxies. The right hand panel, that uses the best estimate for SFR based on the emission of the system, produces the flattest relation between CFE and $\Sigma_\mathrm{SFR}$. Interestingly, the CFE based on the SFR tracers also provides the best agreement with the observations, especially in the starburst regime. Expanding the time interval of Eq. \ref{eq:gamma} to better match the response time scale of the FUV tracer ($>10$ Myr) would in general average out the CFE to lower values. Some uncertainty in the CFE-$\Sigma_\mathrm{SFR}$ relation is introduced in the definition of the $\Sigma_\mathrm{SFR}$ as well. The elongated gas-rich merger has a very irregular shape most of the time, making the exact value of $\Sigma_\mathrm{SFR}$ dependent on image resolution and orientation (as discussed in \citealt{2020ApJ...891....2L}).

Other idealised merger studies such as \citet{2022MNRAS.514..265L}, as well as cosmological simulations such as \citet{2019MNRAS.490.1714P} and \citet{2017ApJ...834...69L}, find similar clear dependence of the CFE on the star formation environment as depicted by the fit by \citet{2010MNRAS.405..857G} and the analytic model by \citet{2012MNRAS.426.3008K} that are shown in Fig. \ref{fig:gamma}. \citet{2017ApJ...834...69L} and \citet{2022MNRAS.514..265L} analyse young clusters ($<50$ Myr and $<50$ Myr, respectively) and find values from a few per cent to almost unity while \citet{2019MNRAS.490.1714P} use a broad range of cluster ages ($<300$ Myr) and only find values up to $\sim 50\%$. The CFE has also been shown to increase towards the central regions of galaxies, where the SFR surface densities, gas densities and mid-plane pressures are higher (\citealt{2017ApJ...834...69L, 2019MNRAS.490.1714P, 2020MNRAS.496..638H, 2022MNRAS.514..265L}; see e.g. \citealt{2013MNRAS.436L..69S} for a similar observational result). Molecular cloud scale simulations including detailed stellar feedback such as protostellar jets by \citet{2021MNRAS.506.3239G} find that CFE correlates with the integrated star formation efficiency (see also \citealt{2019MNRAS.487..364L}) and the GMC surface density.

As a conclusion, the majority of the stellar mass forms in clustered environments, regardless of the detection methodology, as the mass weighted averages are 66$\%$ for the photometric results and $68\%$ for the bound analysis. When the SFR is converted from the dust-corrected FUV emission, the resulting mean value is somewhat lower at $41\%$ due to over-estimated SFR immediately after the starburst. These are somewhat higher values than recovered on average in observations of similar SF surface densities. The isolated nature of our low-mass dwarf merger provides the best formation scenario for SCs as the disruptive tidal forces in the setup are minimal. Inclusion of the cosmological environment, as well as a more detailed modelling of stellar evolutionary processes such as early stellar feedback and non-softened dynamics, might reduce the fraction of stellar mass that survives beyond the first tens or hundreds of Myrs.

\section{Conclusions}

We have taken an observationally motivated look at the star and SC formation process in a simulated merger of two gas rich dwarf galaxies. The simulations have spatial and mass resolution that allow us to resolve the multi-phase structure of the ISM. In the starburst phase the system produces SCs with bound masses of up to $\sim 10^6$ M$_\odot$, that correspond to observed young SSCs. We have post-processed the simulation snapshots using the radiative transfer code \textsc{skirt 9} that we have used to model the SEDs produced by the three-dimensional distribution of individual stars and SCs, while taking into account dust obscuration.  The resulting spectra have been reduced to \textit{HST}-equivalent images at 10 and 50 Mpc galactic distances, corresponding to 1.5 and 7.5 pc per pixel resolution. The integrated and spatially resolved SEDs have then been processed through various commonly used photometric broad-band filters, and \textit{HST}/\textit{JWST} like PSFs and noise levels. With \textit{JWST} the stellar continuum of the most intense star formation region would be observable in 10 000 sec exposures out to redshift $z \lesssim 0.5$. Beyond $z \gtrsim 0.5$ deeper observations or application of magnification by lensing would be required. Inclusion of nebular continuum and line emission (e.g. H$\alpha$ and [OIII]) would also increase the flux in the redshifted UV-visual bands of interest to \textit{JWST}.

The conversion from filter-specific flux to SFR recovers surprisingly well the merger phases and variations of the SFRs when the \mbox{24 $\mu$m} corrected UV tracers are used. This specific combination of tracers provides the SFR to within a factor of two for the majority of the simulation. For comparison, the UV tracers and the IR tracers alone most often underestimate the SFR by up to an order of magnitude for phases of high and low SFR, respectively. The lag from the lifetimes of the massive stars that produce the majority of the UV emission result in an overestimation of the SFR right after the starburst peaks. When compared to observed dwarf starburst galaxies, the SED of our starburst is very similar to what is seen in galaxies of similar mass and star formation rates. The spatial distribution of emission matches the extent and structure of emission in, for example, NGC 1569 that is an actively star-forming dwarf galaxy that also includes young SCs in the SSC mass range. The agreement with observed spectra of nearby dwarf starbursts support the scientific fidelity of the underlying ISM and star formation model.

We have used the photometry at visual bands to identify SCs using a detection pipeline that mimics the production of cluster catalogues in observational surveys of unresolved SCs. We also have at our disposal information of the truly bound cluster population from earlier studies of this specific merger. Especially in images with high resolution, the majority of the bound clusters are captured within the photometrically detected apertures. However, many of the apertures may contain up to tens of bound clusters when the merger is undergoing its starburst phase. In the starburst, we are able to fit power-law mass functions with slopes between $-1.6$ and $-1.9$ to the population of bound clusters contained within the four single brightest apertures of typical aperture size at a 50 Mpc distance. In our simulations SCs form hierachically following power-law mass functions even on small spatial scales of $\sim 50$ pc. This highlights the challenge of observing the formation of individual SSCs due to the hierarchical nature of their formation process. At larger distances (worse spatial resolution, more confusion due to noise) the separation of single objects becomes increasingly difficult. Blending causes the CMF slope to become shallower by $\sim 0.3$ even at a distance of 10 Mpc, as the masses of the lower mass (non-dominant) clusters in each bright aperture are integrated together. This effect is stronger for phases of high star formation rates. At 50 Mpc distance, the mass of the brightest apertures overestimates the respective bound mass by up to a factor of 2.9. We therefore suspect that especially due to the highly hierarchical and obscured nature of the SC formation process, the masses of more distant observed SSCs may indeed be over-estimated. 

Finally, the inclusion or exclusion of short-lived open clusters and unbound associations in the definition of the cluster formation rate can slightly affect the inferred fraction of star formation that occurs in clusters. A larger effect on the efficiency is caused by inaccurate estimates of the total SFR. Based on the various phases of star formation in our merger, we see the largest differences between the photometric analysis (inclusive) and the analysis of pure bound clusters (exclusive) at times of low star formation activity. The hierarchical formation of long-lived (i.e. bound), massive SCs requires more extreme star formation environments, whereas smaller mass and unbound clusters that are able to form even in single collapse events can form in relatively quiescent environments. It is therefore not surprising that we see a slightly weaker environmental dependence of the CFE when we include all (young) cluster-like concentrations of light in the definition of CFR, compared to including only bound young clusters as  in \citet{2020ApJ...891....2L} where the results clearly resembled the analytic model of \citet{2012MNRAS.426.3008K}. Conceptually, these two results correspond to observations that consider strictly young stellar concentrations (regardless of boundness, e.g. \citealt{2017ApJ...849..128C, 2020MNRAS.499.3267A}) and those that emphasise cluster boundness (e.g. including older clusters as well,  \citealt{2008MNRAS.390..759B, 2010MNRAS.405..857G}). The photometric CFE results during relatively more quiescent periods agree qualitatively with observations of young SCs in dwarf galaxies in the ANGST survey, while the starburst periods better agree with more massive actively star-forming galaxies regardless of the analysis method. \citet{2022MNRAS.509.5938H} have concluded that the numerical implementation used in the present study forms SCs that are compact in size and predominantly bound, and therefore not susceptible to disruption processes. Capturing all relevant processes that act on sub-GMC scales, such as proto-stellar feedback and collisional stellar dynamics, might allow for more early cluster disruption. This may in turn have a decreasing effect on the generally high recovered CFE in our simulations especially in quiescent environments, bringing the CFE values closer to observed results for corresponding $\Sigma_\mathrm{SFR}$. However, as we have in general concentrated on young clusters in the analysis of the CFE, the trends between CFE and $\Sigma_\mathrm{SFR}$ should not be significantly affected.

\section*{Acknowledgements}

We would like to thank Angela Adamo for a constructive referee report that helped to improve the data reduction.
We thank Peter Camps for the discussion regarding the optimization of the \textsc{skirt 9} post-processing. NL and TN acknowledge the computing time granted by the LRZ (Leibniz-Rechenzentrum) on SuperMUC-NG under the
project number pn72bu. TN acknowledges support from the Deutsche Forschungsgemeinschaft (DFG, German Research Foundation) under Germany's Excellence Strategy - EXC-2094 - 390783311 from the DFG Cluster of Excellence "ORIGINS". The computations were carried out at CSC -- IT Center for Science Ltd. in Finland and the MPA cluster FREYA hosted by The Max Planck Computing and Data Facility (MPCDF) in Garching, Germany.

This research made use of \textsc{python} packages \textsc{scipy} \citep{2020SciPy-NMeth}, \textsc{numpy} \citep{2020NumPy-Array}, \textsc{matplotlib} \citep{Hunter:2007}, \textsc{pygad} \citep{2020MNRAS.496..152R}, \textsc{h5py} \citep{collette_python_hdf5_2014}, \textsc{pyphot}\footnote{https://mfouesneau.github.io/docs/pyphot/}, \textsc{astropy},\footnote{http://www.astropy.org} a community-developed core \textsc{python} package for Astronomy \citep{2013A&A...558A..33A,2018AJ....156..123A} and \textsc{photutils} that is an \textsc{astropy} package for detection and photometry of astronomical sources \citep{2021zndo...4624996B}.

\section*{Data Availability}

The data will be made available on reasonable request to the corresponding author.



\bibliographystyle{mnras}
\bibliography{references} 






\bsp	
\label{lastpage}
\end{document}